\documentclass[acmsmall,screen,review=false]{acmart}

\usepackage{graphicx}
\usepackage{amsmath}
\usepackage{colortbl}
\usepackage{arydshln}
\usepackage{multirow}
\usepackage{enumerate}
\usepackage{makecell}
\usepackage{amsmath,amsfonts}
\usepackage{algorithm}
\usepackage{algorithmic}
\usepackage{graphicx}
\usepackage{textcomp}
\usepackage{xcolor}
\usepackage{threeparttable}
\usepackage{caption}
\usepackage{hyperref}
\usepackage{varioref}
\usepackage{cleveref}
\usepackage{booktabs}
\usepackage[most]{tcolorbox}
\definecolor{lightgray}{rgb}{249, 249, 249}
\usepackage{adjustbox}

\newtcolorbox{mybox}{
  breakable,
  arc=3mm,
  colback=lightgray,
  colframe=black,
  boxrule=0.3mm,
}

\usepackage{float}
\usepackage{subfig}

\newcommand\jx[1]{\color{black}{#1}}

\usepackage[framemethod=TikZ]{mdframed}
\definecolor{blueish}{RGB}{250, 250, 255}
\definecolor{greenish}{RGB}{200, 255, 200}
\definecolor{redish}{RGB}{255, 200, 200}
\definecolor{highlight}{RGB}{175, 255, 100}
\definecolor{darkred}{RGB}{139, 0, 0}
\definecolor{gray95}{gray}{0.05}
\definecolor{rowgray}{RGB}{224, 224, 224}
\newmdenv[
    tikzsetting= {fill=blueish},
    skipabove=0.33em,
    skipbelow=0.33em,
    linewidth=1pt,
    innerleftmargin=4pt,
    innerrightmargin=4pt,
    innertopmargin=2pt,
    innerbottommargin=2pt,
    linecolor=gray95,
    roundcorner=2pt, 
    shadowsize=4pt,
    shadowcolor=gray95
]{answerbox}
\newenvironment{result}
{\begin{answerbox}}
{\end{answerbox}}
\newcommand{\RS}[2]{
    \begin{result}
        \textbf{Summary of RQ#1:~}{ #2}
    \end{result}
}

\AtBeginDocument{
  }

\setcopyright{acmcopyright}
\copyrightyear{2023}
\acmYear{2023}
\acmDOI{XXXXXXX.XXXXXXX}

\acmConference[Conference acronym 'XX]{Make sure to enter the correct
  conference title from your rights confirmation email}{June 03--05,
  2018}{Woodstock, NY}
\acmPrice{15.00}
\acmISBN{978-1-4503-XXXX-X/18/06}

\begin{document}

\title{CodeScore: Evaluating Code Generation by Learning Code Execution}

\author{Yihong Dong}
\email{dongyh@stu.pku.edu.cn}
\author{Jiazheng Ding}
\email{dingjz@stu.pku.edu.cn}
\author{Xue Jiang}
\email{jiangxue@stu.pku.edu.cn}
\author{Ge Li} \authornote{Corresponding author}
\email{lige@pku.edu.cn}
\author{Zhuo Li}
\email{lizhmq@pku.edu.cn}
\author{Zhi Jin}
\email{zhijin@pku.edu.cn}

\affiliation{
  \institution{Key Laboratory of High Confidence Software Technologies (Peking University), Ministry of Education; School of Computer Science, Peking University, Beijing}
  \country{China}
}

\renewcommand{\shortauthors}{Dong et al.}

\begin{abstract}

A proper code evaluation metric (CEM) profoundly impacts the evolution of code generation, which is an important research field in NLP and software engineering. Prevailing match-based CEMs (e.g., BLEU, Accuracy, and CodeBLEU) suffer from two significant drawbacks. 
1. They primarily measure the surface differences between codes without considering their functional equivalence. However, functional equivalence is pivotal in evaluating the effectiveness of code generation, as different codes can perform identical operations.
2. They are predominantly designed for the Ref-only input format. However, code evaluation necessitates versatility in input formats. Aside from Ref-only, there are NL-only and Ref\&NL formats, which existing match-based CEMs cannot effectively accommodate.
In this paper, we propose CodeScore, a large language model (LLM)-based CEM, which estimates the functional correctness of generated code on three input types. To acquire CodeScore, we present UniCE, a unified code generation learning framework, for LLMs to learn code execution (i.e., learning PassRatio and Executability of generated code) with unified input. Extensive experimental results on multiple code evaluation datasets demonstrate that CodeScore absolutely improves up to 58.87\% correlation with functional correctness compared to other CEMs, achieves state-of-the-art performance, and effectively handles three input formats. 
\end{abstract}

\ccsdesc[500]{Software and its engineering~Software creation and management}
\ccsdesc[500]{Computing methodologies~Artificial intelligence}

\keywords{Code Evaluation, Code Pre-trained Language Model, Code Generation.}

\maketitle

\section{Introduction}
Automatic evaluation of code generation is significant and promising in the fields of natural language processing (NLP) and software engineering. Due to the great potential of code generation in reducing development costs and revolutionizing programming modes, both industry and academia have devoted substantial attention to it \cite{alphacode, MukherjeeWCRCJ21, YinN18, CodeX, Industry, dong2023antecedent}. Code generation has achieved remarkable developments in the past few years \cite{fried2022incoder, codegen, CODEP, jiang2023self, li-etal-2024-deveval}, but CEMs still need to catch up. It is challenging to evaluate the competitiveness of various approaches without proper CEM, which hampers the development of advanced techniques for code generation. A range of code generation subtasks would benefit from valid code evaluation, including code completion \cite{GuoS0DBA22, LuDHGHS22}, code translation \cite{RoziereLCL20, Zhu0R22}, code search \cite{SunFCTHZ22, NS3}, etc. Therefore, research on code evaluation is necessary and should be put on the agenda. 

\begin{figure}[t]
	\centering
    \includegraphics[width=0.85\textwidth]{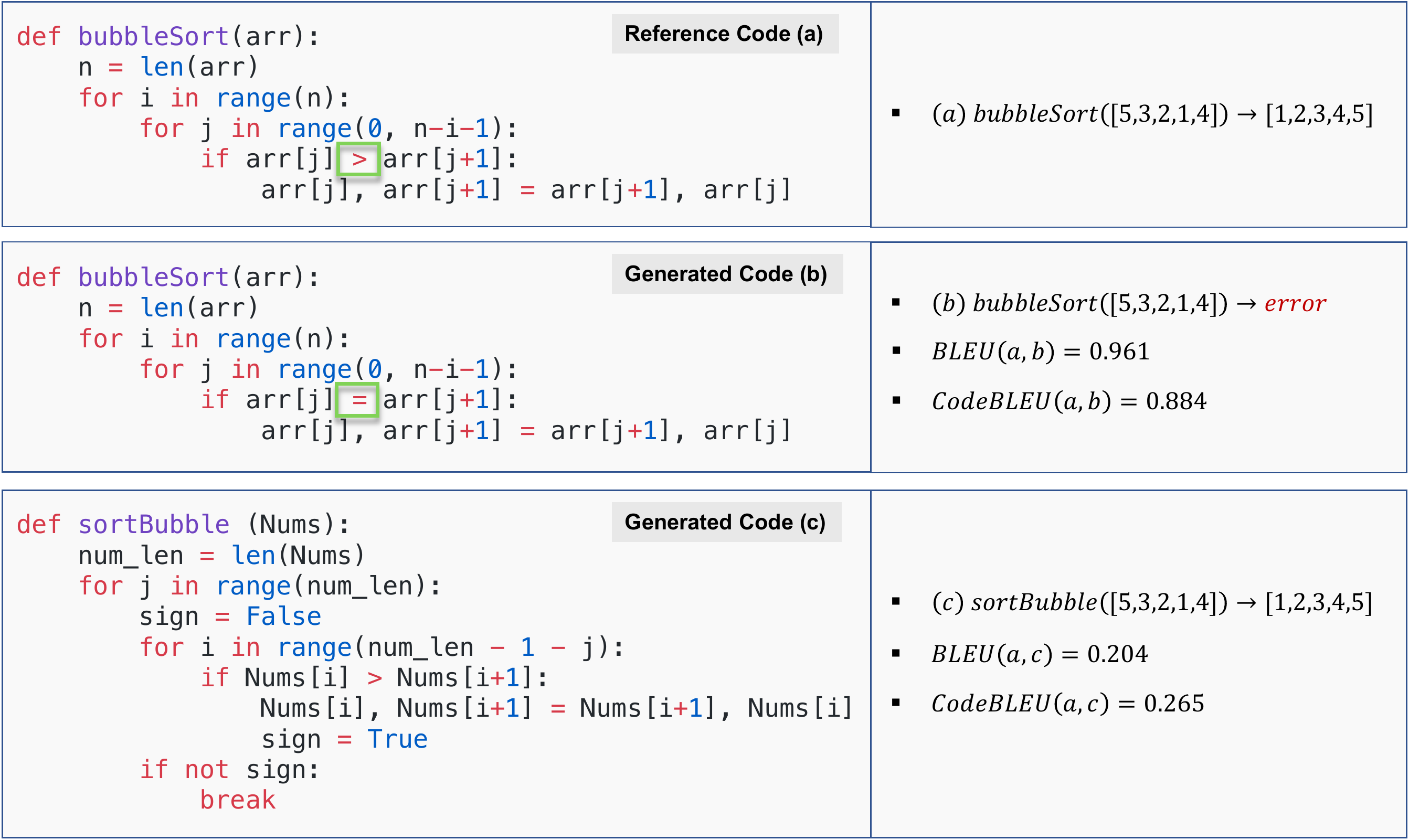}
	\caption{Results of evaluating the generated code implementing bubble sort using different CEMs. BLEU and CodeBLEU score the truly functional correct code (c) lower than the incorrect code (b).}
	\label{code_relationship}
\end{figure}

Some commonly used match-based CEMs treat code as text, such as BLEU \cite{Bleu} and Accuracy, which focus on basic and lexical-level features. They compute scores mainly based on n-gram co-occurrence statistics. CodeBLEU \cite{CodeBLEU} additionally takes into account the structure of code, i.e., abstract syntax tree and data flow. However, the preceding CEMs have deficiencies in identifying code relationships, because code is mainly evaluated based on functional correctness rather than exact/fuzzy match to reference code, and match-based CEMs cannot account for the large and complex space of code functionally equivalent to reference code \cite{dataset_previous_1}. For example, in Fig. \ref{code_relationship}, code (a) and code (b) have a much higher similarity of tokens or structures than code (c). However, through execution, we realize that code (a) and code (c) are different renderings of the same function. By contrast, the execution result of code (b) differs dramatically from both other codes, and code (b) even fails to compile. As a result, merely measuring the similarity of token/structure is insufficient for code evaluation.

LLMs pre-trained on code have demonstrated outstanding results in code generation tasks \cite{CodeX, fried2022incoder, alphacode, dongself, dong2024generalization}\nocite{li2024evocodebench, dong-etal-2024-pace}, which are fundamentally dependent on exceptional code comprehension. Excellent code comprehension is a crucial element for facilitating code evaluation. We hypothesize that LLMs pre-trained on code possess the ability to evaluate code. However, due to the training strategy of predicting the next token according to context, they lack awareness of evaluating code for functional correctness. Our objective is to instruct LLMs to evaluate code effectively in terms of functional correctness.

Another issue that requires resolution is that the existing match-based CEMs are exclusively confined to the Ref-only (consider only reference code) input format. This restriction presents three inherent disadvantages. First, for any code generation task, the correct solutions are not finite, but rather, they are inexhaustible. In this context, the provided reference code merely represents one correct solution among a vast multitude. Therefore, it is overly narrow to compare the generated code solely with one correct solution. Second, they neglect the natural language (NL) description, which is a rich repository of information and a real requirement source. Third, these metrics are unusable in the absence of a reference code. This situation is quite commonplace in real-world evaluations where a correct solution is not always readily available. It is similar to code grading techniques in education, where grading often needs to be flexible and adaptable to different solutions that may not have a single correct answer. 
Therefore, expanding the input format of CEM is necessary.

In this paper, we propose an effective LLM-based CEM, called CodeScore, which measures the functional correctness of generated codes on three input formats (Ref-only, NL-only, and Ref\&NL). To obtain CodeScore, we present a code evaluation learning framework, UniCE, for tuning LLMs to estimate execution similarities with unified input. Specifically, we finetune LLMs to learn PassRatio and Executability of generated code, where Executability is devised to distinguish between compilation errors and output errors for code with PassRatio equal to 0. Generally, codes exhibiting higher functional correctness will pass more test cases, thereby achieving a higher PassRatio \footnote{Note that, although PassRatio varies across different test cases, it tends to yield a higher PassRatio for high-quality code, since we generate a large number of test cases. This phenomenon is somewhat akin to the process of human feedback. Despite the inherent variability in scores assigned by different human evaluators, the overarching trend remains consistent.}. Consequently, for unexecutable codes, the model tends to assign scores approaching zero. In contrast, for codes demonstrating superior functional correctness, the model is likely to assign higher scores.
CodeScore has the following advantages: 1) CodeScore has excellent evaluation performance, which achieves state-of-the-art performance correlation with functional correctness on multiple code evaluation datasets. 2) CodeScore provides three application scenarios (Ref-only, NL-only, and Ref\&NL) for code evaluation with unified input, while traditional CEMs only consider Ref-only. Our major contributions can be summarized as follows: 

\begin{itemize}
    \item We propose an efficient and effective LLM-based CEM, CodeScore, that accommodates the functional correctness of generated codes from an execution viewpoint.\footnote{\url{https://huggingface.co/dz1/CodeScore}}
    \item We present UniCE, a unified code evaluation learning framework based on LLMs with unified input, which assists models in learning code execution and predicting an estimate of execution PassRatio.\footnote{\url{https://github.com/Dingjz/CodeScore}}
    \item We construct three code evaluation datasets based on public benchmark datasets in code generation, called APPS-Eval, MBPP-Eval, and HE-Eval, respectively. Each task of them contains an NL description, several reference codes, 10+ generated codes, and 100+ test cases.\footnote{\url{https://github.com/YihongDong/CodeGenEvaluation}} 
    \item CodeScore substantially outperforms match-based CEMs and LLM-based EMs, and achieves state-of-the-art performance on multiple code evaluation datasets.
\end{itemize}

\section{Background \& Related Work}
\label{Related Work}
In this section, we first introduce code generation, and then discuss code evaluation based on three types of EMs, including Match-based CEMs, Execution-based CEMs, and LLM-based EMs.

\subsection{Code Generation} \jx{
Code generation technology can automatically generate source code for software, achieving the purpose of machine-driven programming based on user requirements. 
Due to the rapid growth of code data and the continuous improvement of deep learning model capabilities, using deep learning for program generation has become the mainstream research direction \citep{RaychevVY14, LingBGHKWS16, YinN18, WeiBolin, SunZXSMZ20, MukherjeeWCRCJ21, zhao2023seq2seq, jiang2024seed}. In recent years, the rise of pre-training techniques has provided new momentum for code generation. For example, studies like CodeT5 \citep{CodeT5} and UniXcoder \citep{UniXcoder} pre-train models for completing code generation tasks. As the number of model parameters increases, researchers have observed the phenomenon of performance emergence in large language models (LLMs).
. LLMs such as AlphaCode \citep{alphacode}, CodeGen \citep{codegen}, WizardCoder \citep{wizardcoder}, ChatGPT \citep{ChatGPT}, CodeGeeX \citep{codegeex}, Starcoder \citep{starcoder}, and CodeLlama \citep{codellama} have demonstrated promising code generation performance. 
Currently, code generation technology and tools have been widely adopted in software development, such as Copilot \citep{CodeX}, significantly enhancing the efficiency of developers.
Assessing the quality of generated code has remained a critical problem in the development of code generation technology, directly influencing its advancement and evolution.
}

\subsection{Code Evaluation}
\paragraph{\textbf{Match-based CEMs.}} Besides these commonly used BLEU \cite{Bleu}, Accuracy, and CodeBLEU \cite{CodeBLEU}, some niche CEMs \cite{chrF} are also applied to code evaluation, e.g., METEOR \cite{METEOR}, ROUGE \cite{ROUGE}, and CrystalBLEU \cite{CrystalBLEU}. However, these aforementioned match-based CEMs merely measure the surface-level differences in code and do not take into account the functional correctness of the generated code. 

\paragraph{\textbf{Execution-based CEMs.}} They attempt to handle these issues by running tests for generated code to verify its functional correctness \cite{Pass@k, APPS, AixBench}. However, they come with several caveats: 1) It assumes that test cases have been given and all dependencies have been resolved. For each code generation task, supplying adequate test cases is a burden in practice, and the dependencies required vary from task to task. 2) Enormous computational overhead needs to be afforded. All generated code requires execution separately for each corresponding test case, which leads to enormous CPU and I/O overhead. 3) Execution with isolation mechanisms. The generated code could have some security risks, such as deleting files on the disk or implanting computer viruses, especially if the training data of code generation models is attacked. In a word, they are usually costly, slow, and insecure, which are often unavailable or ineffective in real-world scenarios. 

\paragraph{\textbf{LLM-based EMs.}}
Effective evaluation of generated results is hard for both text and code generation. They likewise face the same issue of poor evaluation metrics (EMs). A recent popular trend in evaluating text generation is the design of automatic EMs based on LLMs. A part of LLM-based EMs \cite{COMET-21, UniTE, COMET-22} follows COMET \cite{COMET} to learn high-quality human judgments of training data, which is a problem for code evaluation to obtain. Another part relies on LLM extracting token embeddings to calculate scores like BERTScore \cite{BERTScore}, such as \cite{MoverScore, BLEURT, BARTScore, Sentence-BERT}. A concurrent work named CodeBERTScore \cite{CodeBERTScore} tries to use the same way as BERTScore with LLM pre-trained on code. However, they do not teach LLMs to learn code evaluation effectively, in other words, LLMs are still confused about how to evaluate code. Therefore, they exhibit suboptimal performance in code evaluation, as evidenced by our experimental results.

\begin{figure}[t]
	\centering
	\includegraphics[width=0.85\textwidth]{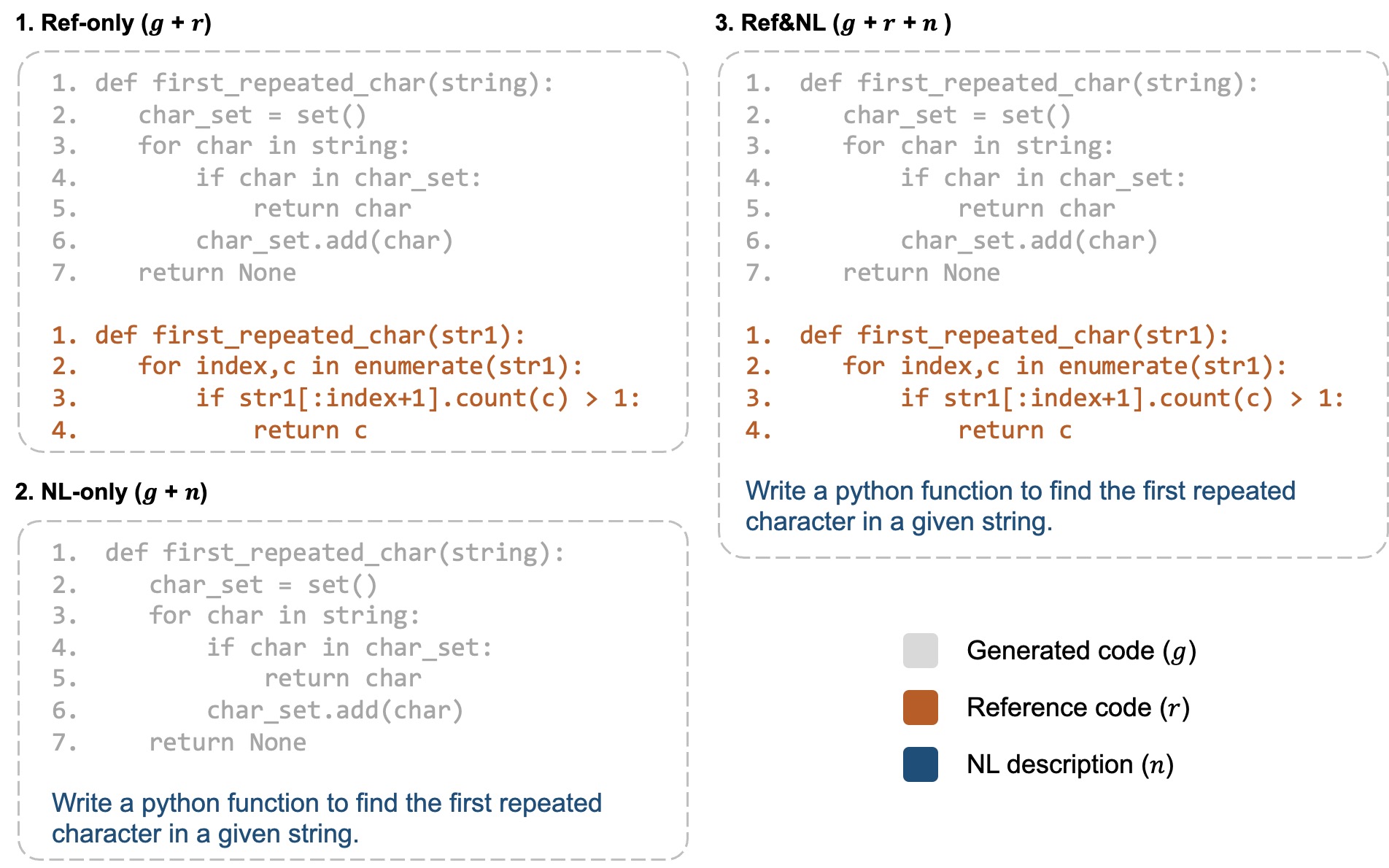}
	\caption{Examples of three input formats for code evaluation.}
	\label{InputType}
\end{figure}

\section{Methodology}
In this section, we first introduce our proposed CEM CodeScore, and then describe a unified code evaluation learning framework (i.e., UniCE), which is used to yield the CodeScore.
\label{Methodology}
\subsection{CodeScore} 

For a code generation task $p \in P$, let the test case set of $p$ as $C_p = \{(\mathcal{I}_{p,c}, \mathcal{O}_{p,c})\}_{c \in C_p}$, a set of paired test case input $\mathcal{I}_{p,c}$ and test case output $\mathcal{O}_{p,c}$. Although the potential program space can be boundless, test cases permit automatic evaluation of code generation capability. Thus, in contrast to most other text generation tasks, human judgment is not always necessary for code generation. 

We measure the functional correctness with \textbf{PassRatio} $\in [0,1]$, which is defined as
\begin{equation}
    \label{PassRatio}
    \operatorname{PassRatio} = \frac{1}{|C_{p}|} \sum_{c\in C_{p}} \mathbb{I}\left\{\operatorname{Eval}\left(\mathbf{g}_p, \mathcal{I}_{p,c} \right)=\mathcal{O}_{p,c}\right\}.
\end{equation}
where $|\cdot|$ indicates the element number of a set, $\mathbb{I}\left\{\cdot\right\}$ is an indicator function, which outputs 1 if the condition is true and 0 otherwise, and $\operatorname{Eval}\left(\mathbf{g}_p, \mathcal{I}_{p,c} \right)$ represents an evaluation function that obtains outputs of code $\mathbf{g}_p$ by way of executing it with $\mathcal{I}_{p,c}$ as input. 

Our framework UniCE can learn existing CEMs, including PassRatio and Passability \footnote{Passability is defined as $\frac{1}{|C_{p}|} \prod_{c\in C_{p}} \mathbb{I}\left\{\operatorname{Eval}\left(\mathbf{g}_p, \mathcal{I}_{p,c} \right)=\mathcal{O}_{p,c}\right\}.$}. In this paper, we choose PassRatio since we want to study execution similarity and continuous PassRatio can better reflect the execution similarity of different codes than binary Passability. In the case of generated code with PassRatio equal to 0, we also use binary \textbf{Executability} to distinguish whether the generated code can be executed successfully with all given test cases, and thus measure its quality.
\begin{equation}
    \operatorname{Executability} = \left\{ 
\begin{aligned}
&1, \ if\ code \ is \ executable,\\
&0, \ otherwise.
\end{aligned}
\right. 
\end{equation}

\begin{figure}[t]
	\centering
	\includegraphics[width=\textwidth]{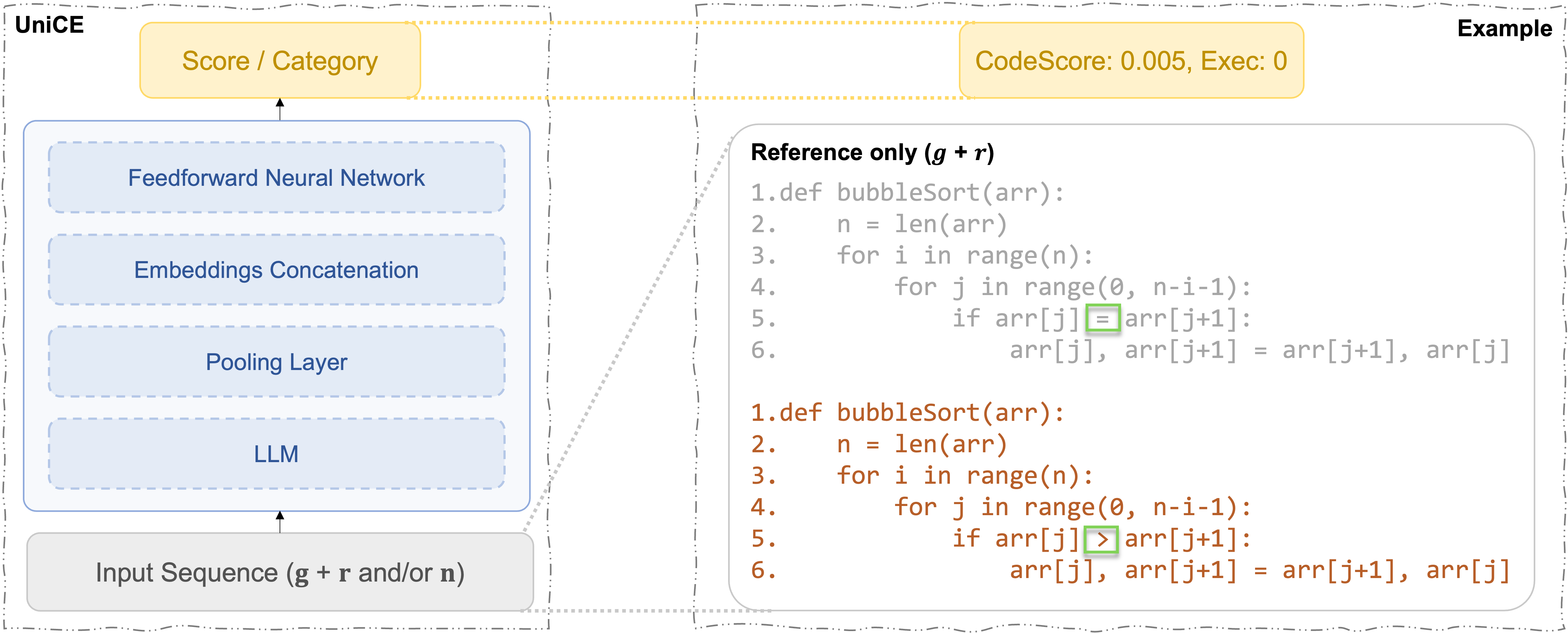}
	\caption{Diagram of UniCE, where the left side of the figure shows its model architecture, and the right side of the figure shows the example (case in Fig. \ref{code_relationship}) of input and output.}
	\label{UniCE}
\end{figure} 

Given a unified input sequence $\mathbf{x}$ that admits the following three types, as shown in Fig. \ref{InputType}: 
\begin{enumerate}[\indent\indent 1.]

\item \textbf{Ref-only} ($\mathbf{g}$ + $\mathbf{r}$): Generated code concatenated with its reference code,

\item \textbf{NL-only} ($\mathbf{g}$ + $\mathbf{n}$): Generated code concatenated with its NL description of requirements,

\item \textbf{Ref\&NL} ($\mathbf{g}$ + $\mathbf{r}$ + $\mathbf{n}$): Generated code concatenated with both its reference code and NL.
\end{enumerate} 

UniCE yields a scalar CodeScore $\in [0,1]$ and a binary number Exec:
\begin{equation} \label{CodeScore, Exec}
    (\operatorname{CodeScore}, \operatorname{Exec}) = \operatorname{UniCE}(\mathbf{x}), 
\end{equation}
where $\operatorname{Exec} = 1$ if $\mathbf{g}$ can be executed successfully with all given test inputs otherwise 0, UniCE is our proposed learning framework, and details of UniCE are presented in Section \ref{UniCESection}. 

We encourage UniCE to learn code execution (i.e., PassRatio and Executability) by minimizing loss function $\mathcal{L}$, which consists of two components: 
\begin{equation}
    \label{L}
    \mathcal{L} = \mathcal{L}_C + \mathcal{L}_E,
\end{equation}
where \(\mathcal{L}_C\) focuses on predicting PassRatio, and \(\mathcal{L}_E\) on predicting code execution correctness. \(\mathcal{L}_C\) and \(\mathcal{L}_E\) are defined as:
\begin{equation}
    \mathcal{L}_C = \left(\operatorname{CodeScore} - \operatorname{PassRatio}\right)^2,
\end{equation}
\begin{equation}
    \mathcal{L}_E = - \log \mathbf{p}(\operatorname{Exec}| \operatorname{Executability}),
\end{equation}
where \(\mathcal{L}_C\) measures the squared difference between the predicted \(\operatorname{CodeScore}\) and the actual PassRatio. \(\mathcal{L}_E\) represents the negative log of the conditional probability of \(\operatorname{Exec}\) given its Executability. The conditional probability is modeled as:
\begin{equation}
    \mathbf{p}(\operatorname{Exec}|\operatorname{Executability}) = \left\{ 
\begin{aligned}
&\mathbf{p}(\operatorname{Exec}), &\text{if } \operatorname{Executability}=1, \\
&1 - \mathbf{p}(\operatorname{Exec}), &\text{otherwise},
\end{aligned}
\right. 
\end{equation}
where \(\mathbf{p}(\operatorname{Exec})\) is the predicted probability of successful execution. 

\subsection{UniCE}
\label{UniCESection}
UniCE relies on LLMs to extract representations of $\mathbf{x}$ and can work with existing pre-trained LLMs. A detailed illustration of the UniCE framework is presented in Fig. \ref{UniCE}.

\subsubsection{Pooling Layer}
For LLMs, the pooling layer plays a critical role in enhancing the model's ability to capture and utilize information more effectively. The work \cite{TenneyDP19, BERTScore, COMET} shows that exploiting information from different layers of LLM generally results in superior performance than only the last layer. Therefore, following the work \cite{ELMO}, we pool information from different layers by using a layer-wise attention mechanism and the final embedding of a token $t$ can be computed as:
\begin{equation}
    e_t = \gamma \sum^l_{k=1} e^k_t h^k,
\end{equation}
where $l$ indicates the number of layers, and $\gamma$ and $h^k$ are trainable weights.

\subsubsection{Unified Embedding}
We require an efficient and comprehensive representation to encapsulate the unified input sequence $x$. Generally, there are two standard methods to extract the representation of $x$, i.e., averaging all token embeddings and using the first token embedding. While the first method is straightforward and includes information from all tokens, it may dilute the significance of more critical tokens and introduce extraneous noise. The first token of our base models is specifically designed to be a summary token\footnote{During the pre-training of our base models (such as CodeBert, GraphCodeBert, and UniXcoder), the first input token is typically the CLS token (short for "classifier"), which enables the model to consider global contextual information during the encoding process through self-supervised learning methods. Therefore, the representation of this first token is usually used to represent the entire input sequence.}.  Moreover, the work \cite{TransQuest, UniTE} also proves the superiority of using the first token embedding compared to averaging all token embeddings in various applications. Thus, we employ the final embedding of the first token $e_{first}$ as the representation of the unified input sequence $x$.

\subsubsection{Unified training}
In UniCE, $e_{first}$ is fed to a feed-forward neural network to output a score and/or a category. To unify three evaluation input formats into UniCE, we apply multi-task learning for training. Specifically, for each step, we assign three sub-steps for three input formats, yielding $\mathcal{L}^{Ref}$, $\mathcal{L}^{NL}$, and $\mathcal{L}^{Ref + NL}$, respectively. A Ref\&NL data can be regarded as three input format data to yield three losses, while Ref-only and NL-only data can only compute the corresponding $\mathcal{L}^{Ref}$ and $\mathcal{L}^{NL}$. The final learning objective of UniCE is to minimize $\mathcal{L}^{Uni}$:  
\begin{equation}
    \mathcal{L}^{Uni} = \mathcal{L}^{Ref} + \mathcal{L}^{NL} + \mathcal{L}^{Ref + NL},
\end{equation}
where $\mathcal{L}^{Ref}$, $\mathcal{L}^{NL}$, and $\mathcal{L}^{Ref + NL}$ are compute via Eq. \ref{L} using corresponding format data as input.

\section{Evaluation}
We aim at answering the following research questions (RQs):
\begin{itemize}
    \item RQ1: What is the performance of CodeScore on code evaluation tasks, compared to other EMs?
    \item RQ2: Can Exec effectively identify whether a generated code can be executed when all dependencies are met?
    \item RQ3: What is the contribution of $L^{Uni}$ to UniCE for three input formats, compared to their respective losses?
    \item RQ4: How reasonable are the evaluations of CodeScore and other EMs from a human perspective?
    \item RQ5: How do CodeScore and other EMs perform on code evaluation tasks in a practical scenario?
\end{itemize}

\jx{Our five RQs aim to evaluate the efficacy and practicality of our approach compared to existing EMs. RQ1 and RQ4 assess our approach against current EMs through experiments and human evaluations, ensuring a comprehensive analysis from both quantitative and qualitative perspectives. RQ2 and RQ3 involve ablation studies to pinpoint the individual and combined impacts of our approach's main components. RQ5 evaluates our approach's real-world applicability through case studies.}

\subsection{Experiment Setup}
\label{Experiment Setup}
In this section, we introduce datasets, baselines, correlation evaluation, and implementation details.

\begin{table*}[ht!]
	\caption{Statistics of datasets (part 1).}	\label{statistics1}
	\centering
        \resizebox{\textwidth}{!}{
	\begin{tabular}{lcccccccccc}
		\toprule
		 \multirow{2}{*}{Dataset} &\multicolumn{3}{c}{Examples Num}& \multicolumn{4}{c}{Avg Num / Task} & \multicolumn{3}{c}{Avg Length}\\
            \cmidrule(r){2-4} \cmidrule(r){5-8} \cmidrule(r){9-11}
		   & Train & Dev & Test & NL & RefCode & GenCode & Extended (Original) TestCase   & NL & RefCode & GenCode  \\
		\midrule
            APPS-Eval  & 267,162 & 33,395 & 33,395 & 1 & 13 & 32 & 181 (13) & 263.8 & 86.3 & 76.8 \\
		MBPP-Eval & 15,679 & 3,000 & 3,000 & 1 & 1 & 24 & 102 (3) & 15.5 & 32.5 & 26.7\\
        HE-Eval & - & - & 4221 & 1 & 1 & 26 & 108 (8) & 61.9 &  24.4 & 41.6 \\ 
		\bottomrule
	\end{tabular}}
\end{table*} 
\subsubsection{Datasets}
We construct three public datasets (named APPS-Eval, MBPP-Eval, and HE-Eval) for code evaluation based on three public benchmark datasets in code generation, i.e., MBPP \cite{MBPP}, APPS \cite{APPS}, and HumanEval \cite{CodeX}. 

\begin{table}[ht!]
	\caption{Statistics of datasets (part 2).}	\label{statistics2}
	\centering
	\begin{tabular}{lcccccc}
		\toprule
		 \multirow{2}{*}{Dataset} &\multicolumn{3}{c}{AvgPassRatio}& \multicolumn{3}{c}{Pass@1}\\
            \cmidrule(r){2-4} \cmidrule(r){5-7}
		   & Train & Dev & Test & Train & Dev & Test \\
		\midrule
            APPS-Eval & 0.3196 & 0.1814 & 0.1790 & 0.0315 & 0.0007 & 0.0011\\
		MBPP-Eval & 0.2832 & 0.2571 & 0.2890 & 0.0674 & 0.0494 & 0.0760\\
            HE-Eval & - & - & 0.3695 & - & - & 0.1591\\
		\bottomrule
	\end{tabular}
\end{table}  

To construct each code evaluation dataset, we first follow primitive NL and reference code in each corresponding base dataset. Then, for each paired NL and reference code in a code evaluation dataset, we generate an average of 20+ codes (generated from various LLMs, including CodeGen 350M\&16B \cite{codegen}, InCoder 1B\&6B \cite{fried2022incoder}, and CodeX 13B\&175B \cite{CodeX}. For HE-Eval dataset, we also consider the latest state-of-the-art LLMs including StarCoder 15.5B \cite{starcoder}, CodeLlama 34B \cite{codellama}, and GPT-4 \cite{GPT-4} besides the aforementioned LLMs.) according to NL and additionally build an average of 100+ correct test cases according to reference code. To obtain these test cases, the following steps were implemented:
\begin{enumerate}[\indent 1)]
    \item Infer the type of input from pre-existing test cases.
    \item Enumerate a collection of inputs constrained by the type of input and task.
    \item Feed the input into the original correct code and get the output by execution (We assume that all external dependencies including third-party libraries have been installed correctly).
\end{enumerate}

Finally, we label each matched NL, reference code, and generated code by executing the generated code with all corresponding test cases to compute PassRatio via Eq. \ref{PassRatio}. Statistics of the datasets are presented in Table \ref{statistics1} and Table \ref{statistics2}\footnote{For each generated code, we employ extended test cases of the corresponding task to compute its PassRatio and Passability. We compute the average number of PassRatio and Passability, i.e., AvgPassRatio and Pass@1, on the train, dev, and test sets of each dataset and display them in Table \ref{statistics2}.}.  \emph{As demonstrated in Table \ref{statistics1} and Table \ref{statistics2}, there are notable disparities in the distributions of NL, RefCode (Reference Code), GenCode (Generated Code), and test cases across the three datasets.} Specifically, 

\begin{itemize}
    \item \textbf{APPS-Eval} has 267,162 training examples and 33,395 examples each for dev and test sets. Each task typically includes 1 NL, 13 RefCode, and 42 GenCode, with average token lengths of 263.8 for NL, 86.3 for RefCode, and 76.8 for GenCode. Extended test cases average 181 per task, compared to the original 13. The AvgPassRatio for train, dev, and test sets are 0.3196, 0.1814, and 0.1790, respectively, while Pass@1 are 0.0315, 0.0007, and 0.0011, respectively.

    \item \textbf{MBPP-Eval} has 15,679 training examples and 3,000 examples each for dev and test sets. Each task typically includes 1 NL, 1 RefCode, and 24 GenCode, with average token lengths of 15.5 for NL, 32.5 for RefCode, and 26.7 for GenCode. Extended test cases average 102 per task, compared to the original 3. The AvgPassRatio for train, dev, and test sets are 0.2832, 0.2571, and 0.2890, respectively, while Pass@1 are 0.0674, 0.0494, and 0.0760, respectively.

    \item \textbf{HE-Eval} has 4,221 test examples. Each task typically includes 1 NL, 1 RefCode, and 26 GenCode, with average token lengths of 61.9 for NL, 24.4 for RefCode, and 41.6 for GenCode. Extended test cases average 108 per task, compared to the original 8. The AvgPassRatio for the test set is 0.3695, while Pass@1 is 0.1591.
\end{itemize}

\subsubsection{Baselines}
We select typical match-based CEMs, LLM-based EMs, and execution-based CEMs as baselines. We present each type of EMs as below.

\emph{\textbf{Match-based CEMs}} include BLEU \cite{Bleu}, Exact Matching Accuracy (Accuracy), CodeBLEU \cite{CodeBLEU}, and CrystalBLEU \cite{CrystalBLEU}, specifically:

\begin{itemize}
    \item \textbf{BLEU} \cite{Bleu} is calculated based on n-gram, and the fluency and correctness of generated code are expressed by calculating the proportion of n consecutive tokens in the correct code, where n is usually set to 4 (i.e., BLEU-4). Considering that shorter codes usually have higher BLEU values, a penalty item is introduced to BLEU as:
\begin{equation*}
    \operatorname{BLEU} =BP \cdot \exp \left(\sum_{m=1}^{n} \omega_{m} \log p_{m}\right),
\end{equation*}
\begin{equation*}
\begin{array}{c}
BP =\left\{\begin{array}{cc}
1, & l_g \geq l_r \\
e^{\left\{1-\frac{r}{l_g}\right\}}, & l_g<l_r
\end{array}\right.
\end{array},
\end{equation*}
where $BP$ represents the penalty item, $l_g$ represents the length of generated code, $l_r$ represents the length of reference code, and $\omega_{m}$ and $p_{m}$ represents the weighted coefficient and precision of $m$-gram, respectively. 

\item \textbf{Accuracy} indicates the percentage of exact matches between generated code and reference code.

\item \textbf{CodeBLEU} \cite{CodeBLEU} additionally takes into account the structure of code, which absorbs the advantages of BLEU in n-gram matching, and further injects code syntax through abstract syntax tree and code semantics through data flow.
\begin{align*}
    \text { CodeBLEU }& = \alpha \cdot \operatorname{BLEU} + \beta \cdot \operatorname{BLEU}_{w e i g h t}\\  & + \delta \cdot \text{Match}_{ast} + \zeta \cdot \text{Match}_{df},
\end{align*}
where $\alpha, \beta, \delta$ and $\zeta$ are weights (usually set to 0.25, as well as in this paper), $\operatorname{BLEU}_{weight}$ is a weighted BLEU with different weights for various tokens, $\text{Match}_{ast}$ is syntactic AST matching, which explores the syntactic information of the code, and $\text{Match}_{df}$ is semantic dataflow matching, which considers the semantic similarity between generated code and reference code.

\item \textbf{CrystalBLEU} \cite{CrystalBLEU} is a metric that calculates BLEU by reducing the noise caused by trivially shared n-grams, such as `(' and `,'.
\end{itemize}

\textbf{LLM-based EMs} contain two well-known and widely used text EMs (BERTScore \cite{BERTScore} and COMET \cite{COMET}) and a concurrent work (CodeBERTScore \cite{CodeBERTScore}), specifically:

\begin{itemize}
    \item \textbf{BERTScore} \cite{BERTScore} is an automatic evaluation metric for text generation, which computes a similarity score for
each token in the generated sentence with each token in the reference sentence with contextual embeddings of BERT \cite{BERT}.
\begin{align*}
        &R_{\mathrm{BERT}}=\frac{1}{|\mathbf{x}|} \sum_{\mathbf{x}_{i} \in \mathbf{x}} \max _{\hat{\mathbf{x}}_{j} \in \hat{\mathbf{x}}} \mathbf{x}_{i}^{\top} \hat{\mathbf{x}}_{j}, \quad P_{\mathrm{BERT}}=\frac{1}{|\hat{\mathbf{x}}|} \sum_{\hat{\mathbf{x}}_{j} \in \hat{\mathbf{x}}} \max _{\mathbf{x}_{i} \in \mathbf{x}} \mathbf{x}_{i}^{\top} \hat{\mathbf{x}}_{j},\\
        &\qquad\qquad\qquad\quad F_{\mathrm{BERT}}=2 \frac{P_{\mathrm{BERT}} \cdot R_{\mathrm{BERT}}}{P_{\mathrm{BERT}}+R_{\mathrm{BERT}}}.
\end{align*}
Following the setting in \cite{BERTScore}, we compute BERTScore with inverse document frequency computed from test sets. 

\item \textbf{COMET} \cite{COMET} provides a text EM by learning human judgments of training data, which leverages cross-lingual pre-trained language modeling to predict the quality of generated text more accurately.

\item \textbf{CodeBERTScore} \cite{CodeBERTScore} is a concurrent work that tries to use the same way as BERTScore with LLM pre-trained on code.
\end{itemize}

\textbf{Execution-based CEM} refers to AvgPassRatio \cite{APPS}.

\begin{itemize}
    \item \textbf{AvgPassRatio} \cite{APPS} is defined as the average proportion of test cases that generated codes $\mathbf{g}_p's$ pass: 
\begin{equation}
    \label{AvgPassRatio}
    \operatorname{AvgPassRatio} = \frac{1}{|P|} \sum_{p\in P} \frac{1}{|C_{p}|} \sum_{c\in C_{p}} \mathbb{I}\left\{\operatorname{Eval}\left(\mathbf{g}_p, \mathcal{I}_{p,c} \right)=\mathcal{O}_{p,c}\right\},
\end{equation}
where $|\cdot|$ indicates the element number of a set, $\mathbb{I}(\cdot)$ is an indicator function, which outputs 1 if the condition is true and 0 otherwise, and $\operatorname{Eval}\left(\mathbf{g}_p, \mathcal{I}_{p,c} \right)$ represents an evaluation function that obtains outputs of code $\mathbf{g}_p$ by way of executing it with $\mathcal{I}_{p,c}$ as input. 
\end{itemize}

As mentioned above, continuous PassRatio (the item of AvgPassRatio) can better reflect the execution similarity of different codes than binary Passability (the item of Pass@1 \footnote{Pass@1 \cite{Pass@k} is defined as the percentage of $\mathbf{ g}_p's$ that pass all test cases of the corresponding $p$: $\frac{1}{|P|} \sum_{p\in P} \frac{1}{| C_{p}|} \prod_{c\in C_{p}} \mathbb{I}\left\{\operatorname{Eval}\left(\mathbf{g}_p, \mathcal{I}_{p, c} \right)=\mathcal{O}_{p,c}\right\},$ where Pass@1 is a more stringent CEM, also known as Strict Accuracy.}). Therefore, in this paper, we mainly compare the correlation between CodeScore and AvgPassRatio in Execution-based CEMs. 

The input format of the proceeding baselines is Ref-only and each of them except COMET is in the range of 0 to 1. 

\begin{table*}[ht!]
	\caption{Correlation comparison of functional correctness on APPS-Eval dataset.}	\label{APPS_Experiments}
	\centering
        \setlength\tabcolsep{2pt}
	\resizebox{0.97\textwidth}{!}{
	\begin{tabular}{lcccccc}
		\toprule
		  Method & Value & $\tau \uparrow$ &  $r_s \uparrow$ & $r_p \uparrow$ & MAE $\downarrow$  & Execution Time $\downarrow$\\
		\midrule
            \textbf{Match-based CEM}\\
            BLEU \cite{Bleu} & 0.0094 & 0.1055 & 0.1156 & 0.0959  & 0.1164 & 1.0 $\times$ (26.0s)\\
            Accuracy & 0.0001 & 0.0079 & 0.0095 & 0.0196 & - & 0.1 $\times$\\
            CodeBLEU \cite{CodeBLEU} & 0.2337 & 0.1035 & 0.1533 & 0.1085 & 0.2005 & 7.8 $\times$ \\
            CrystalBLEU \cite{CrystalBLEU}  & 0.0242 & 0.0906 & 0.1347 & 0.0887 & 0.1709 & 0.3 $\times$ \\
            \midrule
            \textbf{LLM-based EM}\\
            BERTScore \cite{BERTScore}  & 0.8629 & 0.0916 & 0.1375 & 0.0718 & 0.6874 & 56.7 $\times$\\
            COMET \cite{COMET}  & 0.0165 & 0.0904 & 0.1126 & 0.1187 & 0.1751 & 84.0 $\times$ \\
            CodeBERTScore \cite{CodeBERTScore} & 0.7583 & 0.1219 &  0.1801 & 0.1323 &  0.5885 & 27.8 $\times$\\
            \midrule
            \textbf{CodeScore}\\
            \textbf{Ref-only ($\mathbf{g}$ + $\mathbf{r}$)}  \\
            UniCE with $\mathcal{L}^{Ref}$ & 0.1996 & 0.4760 & 0.6473 & 0.6620 & 0.1202 & \multirow{2}{*}{33.7 $\times$} \\
            UniCE with $\mathcal{L}^{Uni}$ & 0.1977 & 0.5033 & 0.6693 & 0.6929 & 0.1128 & \\
            \hdashline 
            \textbf{NL-only ($\mathbf{g}$ + $\mathbf{n}$)} \\
            UniCE with $\mathcal{L}^{NL}$ & 0.2035 & 0.4679 & 0.6359 & 0.6855 & 0.1189 &  \multirow{2}{*}{37.9 $\times$} \\
            UniCE with $\mathcal{L}^{Uni}$ & 0.2016 & 0.4901 & 0.6486 & 0.6905 & 0.1120 &  \\
            \hdashline 
            \textbf{Ref\&NL ($\mathbf{g}$ + $\mathbf{r}$ + $\mathbf{n}$)} \\
            UniCE with $\mathcal{L}^{Ref+NL}$ & 0.1837 & 0.3865 & 0.5419 & 0.6152 & 0.1274 & \multirow{2}{*}{44.2 $\times$}\\
            UniCE with $\mathcal{L}^{Uni}$ & 0.1820 & $\textbf{0.5275} \ \ (\textcolor{red}{\uparrow 40.56\%})$  & $\textbf{0.7040} \ \ (\textcolor{red}{\uparrow 55.07\%})$ & $\textbf{0.7210} \ \ (\textcolor{red}{\uparrow 58.87\%})$ & \textbf{0.1044} &  \\
            \midrule
            \textbf{Execution-based CEM}\\
            13 test cases per task & 0.0978 & 0.3360 & 0.4108 & 0.4987 & 0.1327 & 1.5k $\times$\\
            181 test cases per task  & 0.1790 & - & - & - & - & 20.7k $\times$ \\
		\bottomrule
	\end{tabular}}
\end{table*}

\subsubsection{Correlation Evaluation}
We use three major correlation coefficients in statistics $($i.e., Kendall-Tau${(\tau)}$, Spearman R $(r_s)$, and Pearson R $(r_p)$ to evaluate the correlation between each EM and functional correctness. Furthermore, we use Mean Absolute Error (MAE) to assess the absolute error between them.

\begin{itemize}
    \item \textbf{Kendall-Tau} $\mathbf{(\tau)}$ \cite{kendall1938new} is a statistic used to measure the ordinal association between two measured data:
    \begin{equation}
        \tau=\frac{Concordant - Discordant}{Concordant + Discordant},
    \end{equation}
    where $Concordant$ indicates the number of occurrences that two evaluation data $M^1$ and $M^2$ exist either both $M^1_{i}>M^1_{j}$ and $M^2_{i}>M^2_{j}$ or both $M^1_{i}<M^1_{j}$ and $M^2_{i}<M^2_{j}$, and $Discordant$ indicates the number of occurrences opposite to $Concordant$.

    \item \textbf{Spearman R} $\mathbf{(r_s)}$ \cite{mood1950introduction} is a nonparametric measure of rank correlation (statistical dependence between the rankings of two data):
\begin{equation}
    r_{s}=\frac{\operatorname{cov}(\mathrm{R}(M^1), \mathrm{R}(M^2))}{\sigma_{\mathrm{R}(M^1)} \sigma_{\mathrm{R}(M^2)}},
\end{equation}
where $\operatorname {R} (M^1)$ and $\operatorname {R} (M^2)$ represent the rankings of $M^1$ and $M^2$, ${\displaystyle \operatorname {cov}(\cdot, \cdot)}$ means the covariance function, and ${\displaystyle \sigma_{M}}$ means the standard deviation of $M$.

\item \textbf{Pearson R} $\mathbf{(r_p)}$ \cite{bravais1844analyse} is a measure of linear correlation between two data:
\begin{equation}
    r_{s}=\frac{\operatorname{cov}(M^1, M^2)}{\sigma_{M^1} \sigma_{M^2}}.
\end{equation}

\item \textbf{Mean Absolute Error (MAE)} is a measure of errors between paired data:
\begin{equation}
    \mathrm{MAE}=\frac{\sum_{i=1}^{N}\left|M^1_{i}-M^2_{i}\right|}{N},
\end{equation}
where $|\cdot|$ means the absolute-value function.
\end{itemize}

\begin{table}[ht!]
	\caption{Correlation comparison of functional correctness on MBPP-Eval and HE-Eval datasets.}	\label{HE-Eval_Experiments}
 \setlength\tabcolsep{2pt}
	\centering
	\resizebox{0.97\textwidth}{!}{
	\begin{tabular}{lcccccc}
		\toprule
             \multirow{2}{*}{Method} &\multicolumn{3}{c}{MBPP-Eval} 
             &\multicolumn{3}{c}{HE-Eval} \\
            \cmidrule(r){2-4} \cmidrule(r){5-7} 
             & Value & $r_s \uparrow$ & Execution Time $\downarrow$  & Value & $r_s \uparrow$ & Execution Time $\downarrow$ \\
		\midrule
            \textbf{Match-based CEM}\\
            BLEU \cite{Bleu}  & 0.1186 & 0.1784 & 1.0 $\times$ (0.87s) & 0.2436 & 0.0987 & 1.0 $\times$ (1.96s)\\
            Accuracy & 0.0004 & 0.0299 & 0.1 $\times$ & 0.0011 & 0.0456 & 0.1 $\times$ \\
            CodeBLEU \cite{CodeBERT} & 0.1827 & 0.2902 & 5.0 $\times$ & 0.3452 & 0.3308 & 6.3 $\times$\\
            CrystalBLEU \cite{CrystalBLEU}  & 0.0295 & 0.1645 & 0.3 $\times$ & 0.0427 & 0.2171 & 0.4 $\times$ \\
            \midrule
            \textbf{LLM-based EM}\\
            BERTScore \cite{BERTScore} & 0.8842 & 0.1522 & 62.0 $\times$ & 0.9008 & 0.1214 & 57.5$\times$\\
            COMET \cite{COMET} & -0.5001  & 0.2681 & 69.0 $\times$& 0.0879 &  0.1437 & 58.2$\times$\\
            CodeBERTScore \cite{CodeBERTScore} & 0.7863 &  0.2490 & 44.9 $\times$& 0.8091 & 0.3196  & 47.4 $\times$\\
            \midrule
            \textbf{CodeScore}\\
            Ref-only ($\mathbf{g}$ + $\mathbf{r}$) \\
            UniCE with $\mathcal{L}^{Ref}$ & 0.2975 & 0.5864 & \multirow{2}{*}{17.2 $\times$} & 0.3426 & 0.5671 & \multirow{2}{*}{30.2$\times$}\\
            UniCE with $\mathcal{L}^{Uni}$  & 0.3253 & 0.5999 & & 0.4257 & 0.6378 & \\
            \hdashline 
            NL-only ($\mathbf{g}$ + $\mathbf{n}$) \\
            UniCE with $\mathcal{L}^{NL}$ & 0.3364 & 0.4492 & \multirow{2}{*}{12.6 $\times$} & 0.4985 & 0.5634 & \multirow{2}{*}{30.6$\times$}\\
            UniCE with $\mathcal{L}^{Uni}$ & 0.3327 & 0.5719 & & 0.5624 & 0.6215 & \\
            \hdashline 
            Ref\&NL ($\mathbf{g}$ + $\mathbf{r}$ + $\mathbf{n}$) \\
            UniCE with $\mathcal{L}^{Ref+NL}$ & 0.2905 & 0.5926 & \multirow{2}{*}{20.7 $\times$} & 0.4059 & 0.5965 & \multirow{2}{*}{32.9$\times$}\\
            UniCE with $\mathcal{L}^{Uni}$ & 0.3247 & $ \textbf{0.6027} \ \ (\textcolor{red}{\uparrow 31.25\%})$ & & 0.4731 & $\textbf{0.6597} \ \ (\textcolor{red}{\uparrow 32.89\%})$ & \\
            \midrule
            \textbf{Execution-based CEM}\\
            8 test cases per task & 0.2670 & 0.6826 & 1.0k $\times$ & 0.5994 & 0.6981 & 1.9k $\times$\\
            108 test cases per task  & 0.2890 & - & 28.7k $\times$  & 0.3695 & - & 21.7k $\times$ \\
		\bottomrule
	\end{tabular}}
\end{table}  

\subsubsection{Implementation Details}
In this paper, UniXcoder \cite{UniXcoder} is employed as the base LLM of UniCE, which has the similar parameter size of LLMs in BERTScore \cite{BERTScore} and COMET \cite{COMET}, and larger LLMs can usually lead to better results. 
We format the input sequences as ``[CLS] $g$ [SEP] $r$ [SEP] $n$ [SEP]'', where [CLS] and [SEP] are the special tokens in vocabulary, and we replace $g$, $r$, and $n$ with the generated code, reference code, and NL description, respectively. For the balance of three input formats during the training process, we first sample an NL along with its corresponding generated code and reference code. They are then employed to construct data in three formats: Ref-only, NL-only, and Ref\&NL. Finally, these formats are combined for training UniCE. In all experiments of this paper, we train UniCE on the train set of APPS-Eval. We fine-tune UniCE on the train set of MBPP-Eval only when we specially mention it in our paper. We train UniCE with Adam \cite{adam} optimizer on a single GPU of Tesla A100-PCIe-40G. Empirically, the learning rate is set to $0.001$ and the training epoch is set to 5. The feedforward neural network of UniCE consists of 3 linear transitions with the hyperbolic tangent (Tanh) activation functions, where the corresponding output dimensions are 3,072, 1,024, and 2, respectively.
The input token length is limited to 1024. To mitigate the instability of model training, we exhibit the average performance of UniCE running five times.

\subsection{Experimental Results}
\label{Experimental Results}

\subsubsection{RQ1: Effect of CodeScore}

As illustrated in Table \ref{APPS_Experiments}, CodeScore exhibits a significantly stronger correlation with functional correctness than existing match-based CEMs and LLM-based EMs, which display weak or extremely weak correlations with Ground Truth on APPS-Eval. Compared with the top-performing EM among other EMs, CodeScore achieved absolute improvements of 40.56\%, 55.07\%, and 58.87\% on $\tau$, $r_s$, and $r_p$, respectively. With an $r_s$ value greater than $0.6$, it is evident that there is a strong correlation between CodeScore and Ground Truth. Furthermore, CodeScore has the lowest MAE compared to other EMs. The execution time of CodeScore is similar to other LLM-based EMs and slightly longer than existing Match-based CEMs. However, compared to the $20.7k \times$, $28.7k \times$, and $22.1k \times$ execution time of execution-based CEMs in three code evaluation datasets, CodeScore reduces execution time by three orders of magnitude. We also find that computing execution-based CEMs for code evaluation with a small number of original test cases is insufficient. They have a significant reduction in correlation coefficients compared to using larger extended test cases. In cases where test cases are rare or low-quality, such as on APPS-Eval, the correlation between our CodeScore and Ground Truth even far exceeds that of execution-based CEMs.

We also sought to determine the generalizability. In Table \ref{HE-Eval_Experiments}, we utilize CodeScore, trained on APPS-Eval, to evaluate the code in MBPP-Eval and HE-Eval with fine-tuning and zero-shot settings, respectively. It is important to note that the distributions of NL, RefCode, GenCode, and test cases across these three datasets are quite different\footnote{The average length of NL, RefCode, and GenCode across these three datasets are quite different. The average length of NL in APPS-Eval is 263.8, which far exceeds MBPP-Eval (15.5) and HE-Eval (61.9). The trend of the average length of RefCode and GenCode is similar to NL. For the Average Number of test cases per task, APPS-Eval is extended from 13 to 181, while MBPP-Eval and HE-Eval are extended from 3 and 8 to 102 and 108 respectively.}, as evidenced by their respective statistics shown in Table \ref{statistics1} and Table \ref{statistics2}. Table \ref{HE-Eval_Experiments} reveals the effectiveness of CodeScore on MBPP-Eval and HE-Eval. Remarkably, CodeScore continues to achieve the best correlation compared to other match-based CEMs and LLM-based EMs in these two settings. 

\begin{table*}[ht!]
	\caption{Correlation comparison of functional correctness with different base models on HE-Eval dataset.}	\label{DifferentBaseModels}
	\centering
        \setlength\tabcolsep{8pt}
    \resizebox{\textwidth}{!}{
	\begin{tabular}{lcccccc}
		\toprule
		  Method & Value & $\tau \uparrow$ &  $r_s \uparrow$ & $r_p \uparrow$ & MAE $\downarrow$  & Execution Time $\downarrow$\\
		\midrule
            \textbf{CodeScore} (UniCE with $\mathcal{L}^{Uni}$)\\
            UniXcoder & 0.4731 & 0.4997 & 0.6597 & 0.6486 & 0.2179 & 1.00 $\times$ \\ 
            CodeBert & 0.4809 & 0.4675 & 0.6236 & 0.5622 & 0.2344 & 0.98 $\times$\\ 
            CodeGraphBert & 0.4597 & 0.5073 & 0.6728 & 0.6480 & 0.2281 & 1.07 $\times$ \\
		\bottomrule
	\end{tabular}}
\end{table*}

We conduct the experiments of UniCE based on different code pre-trained models, including CodeBert, CodeGraphBert, and UniXcoder. The results of the experiments are presented in Table \ref{DifferentBaseModels}. We did not observe obvious biases when choosing different base models. One trend we observed is that the better the model's ability to understand the code, the more accurate it is in evaluating the code.

Another intriguing finding is that the quality of CodeBLEU inversely correlates with code length. In other words, the longer code, the poorer correlation between CodeBLEU and Ground Truth. This is likely due to the fact that longer codes tend to incorporate more variations in their syntactic structure. Therefore, for longer codes, the evaluation effect of CodeBLEU gradually degrades to BLEU.

\vspace{6pt}
\RS{1}{CodeScore outperforms match-based CEMs and LLM-based EMs in terms of correlation with functional correctness, even on datasets that it was not trained on. Moreover, CodeScore operates at a speed three orders of magnitude faster than execution-based CEMs.}

\begin{figure}[ht!]
    \centering
    \includegraphics[width=0.7\textwidth]{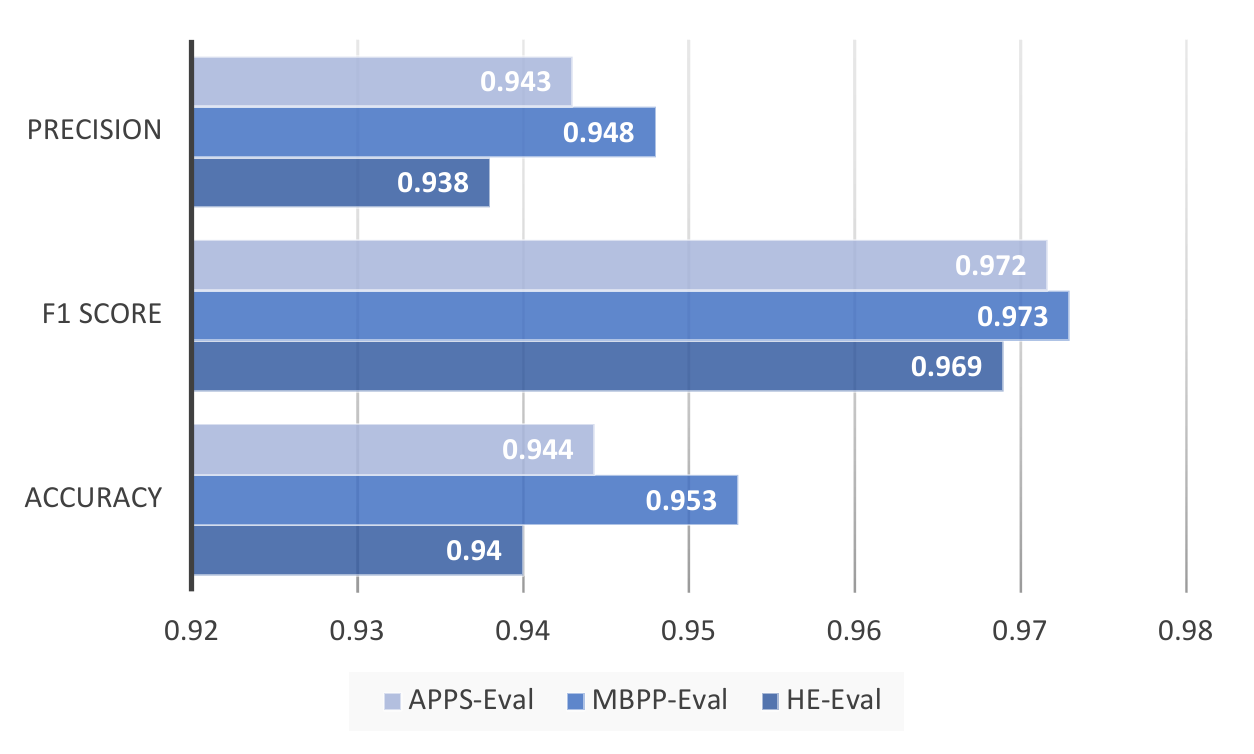}
    \caption{The performance of Exec on APPS-Eval, MBPP-Eval, and HE-Eval datasets.}
    \label{fig:exec}
\end{figure}

\subsubsection{RQ2: Effect of Exec} We also evaluate the performance of Exec on APPS-Eval, MBPP-Eval, and HE-Eval datasets, as shown in Fig. \ref{fig:exec}. The experimental results indicate that Exec demonstrates remarkably high performance in terms of Precision, F1 Score, and Accuracy. Through a comprehensive analysis of all datasets, we find that our approach's performance on the APPS-Eval dataset is inferior to that on the MBPP-Eval dataset. This discrepancy is primarily due to the higher complexity and length of problems in the APPS-Eval dataset compared to those in MBPP-Eval. Furthermore, the performance on the HE-Eval dataset is the poorest, because our approach has not been trained on this dataset. Nevertheless, our approach's performance across various metrics on the HE-Eval dataset exceeded 90\% in the zero-shot setting, indicating its effective transferability to unseen datasets. These results prove that using UniCE to learn code execution is effective for code evaluation.

\vspace{6pt}
\RS{2}{The Exec component in our approach demonstrates extremely high Precision/F1 Score/Accuracy in determining whether the code can be executed when all dependencies are met.}

\subsubsection{RQ3: Effect of $\mathcal{L}^{Uni}$}
As observed from Tables \ref{APPS_Experiments} and \ref{HE-Eval_Experiments}, our proposed $\mathcal{L}^{Uni}$ demonstrates enhancements across all input formats when compared to their respective losses on APPS-Eval, MBPP-Eval, and HE-Eval datasets. With changes in the input format, both the correlation coefficients and MAE between CodeScore and Ground Truth also vary. Generally, the Ref\&NL input format yields superior results, which shows that accommodating NL has a positive effect on evaluating the generated code, while the traditional Ref-only input format omits the valuable information in NL.  Additionally, according to the Avg Length data presented in Table \ref{statistics1}, we discovered that the execution time of CodeScore exhibits a linear, positive relationship with the input length. Regardless of the input formats, our proposed CodeScore provides a commendable evaluation of generated code. This is attributable to the fact that $\mathcal{L}^{Uni}$ aids in training a code evaluation model with a unified input.

\vspace{6pt}
\RS{3}{The component $\mathcal{L}^{Uni}$ in our approach shows positive effects across different input formats.}

\begin{table}[h]
    \caption{Human evaluation for functional correctness.}	 
            \label{HECS}
        \centering
        \begin{tabular}{lc}
            \toprule
              EM & Reasonableness of Evaluation\\
            \midrule
                BERTScore \cite{BERTScore} & 1.3 ± 0.4\\
                CodeBLEU \cite{CodeBLEU} & 2.1 ± 0.5\\
                CodeBERTScore \cite{CodeBERTScore} & 2.2 ± 0.7\\
                CodeScore  & \textbf{3.4} \ \textcolor{red}{($\uparrow$ 54.6\%)} ± 0.3\\
                \midrule
                Ground Truth & 4.6 ± 0.2\\
            \bottomrule
        \end{tabular}
    \end{table}

\subsubsection{RQ4: Human Evaluation}
\label{Human Evaluation}
In this section, we conduct a human evaluation to gauge the validity of our CodeScore. Considering the costliness of human evaluation, we select only five representative EMs for this task, namely, CodeScore, CodeBLEU, BERTScore, CodeBERTScore, and Ground Truth (i.e., PassRatio). All of these EMs are continuous and range from 0 to 1. In accordance with previous work \cite{AixBench} and our experimental setup, we manually assess the validity of each EM in gauging the functional correctness of the generated code. The score for this evaluation is an integer ranging from 0 to 5, where 0 denotes poor and 5 signifies excellent performance.

The human evaluation is conducted on the Python dataset HE-Eval. We randomly select 100 samples \footnote{\jx{Considering the workload of the evaluators, we choose a moderate sample size of 100. Too many samples would exceed the evaluators' capacity.}} from this dataset, each consisting of natural language descriptions, reference code, and generated code. These samples are scored using five EMs, resulting in a total of 100*5 data pairs. We invite ten computer science PhD students, each with over three years of experience in Python development, to serve as evaluators. The 500 code snippets are divided into 10 groups, with each questionnaire containing one group. We randomly list the generated code with reference code and NL and the corresponding EM score on the questionnaire. Each group is evaluated anonymously by one evaluator, and the final score is the average of all evaluators’ scores. Evaluators are allowed to search the Internet for unfamiliar concepts.

\begin{figure}[h!]
	\centering
	\subfloat[Case I]{\label{fig:a}\includegraphics[width=\textwidth]{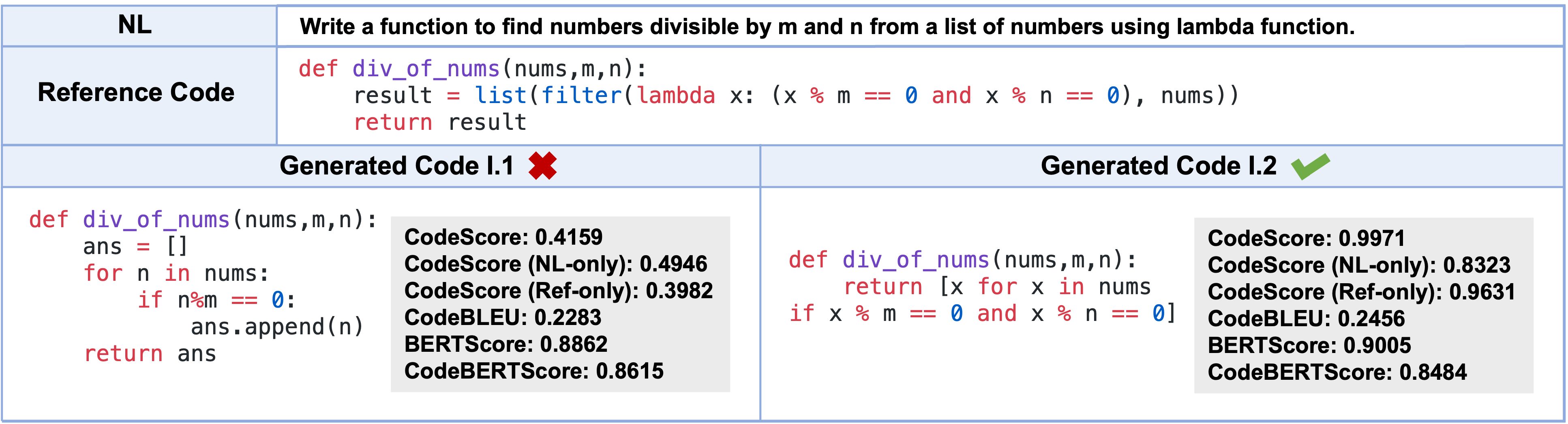}}\\
	\subfloat[Case II]{\label{fig:b}\includegraphics[width=\textwidth]{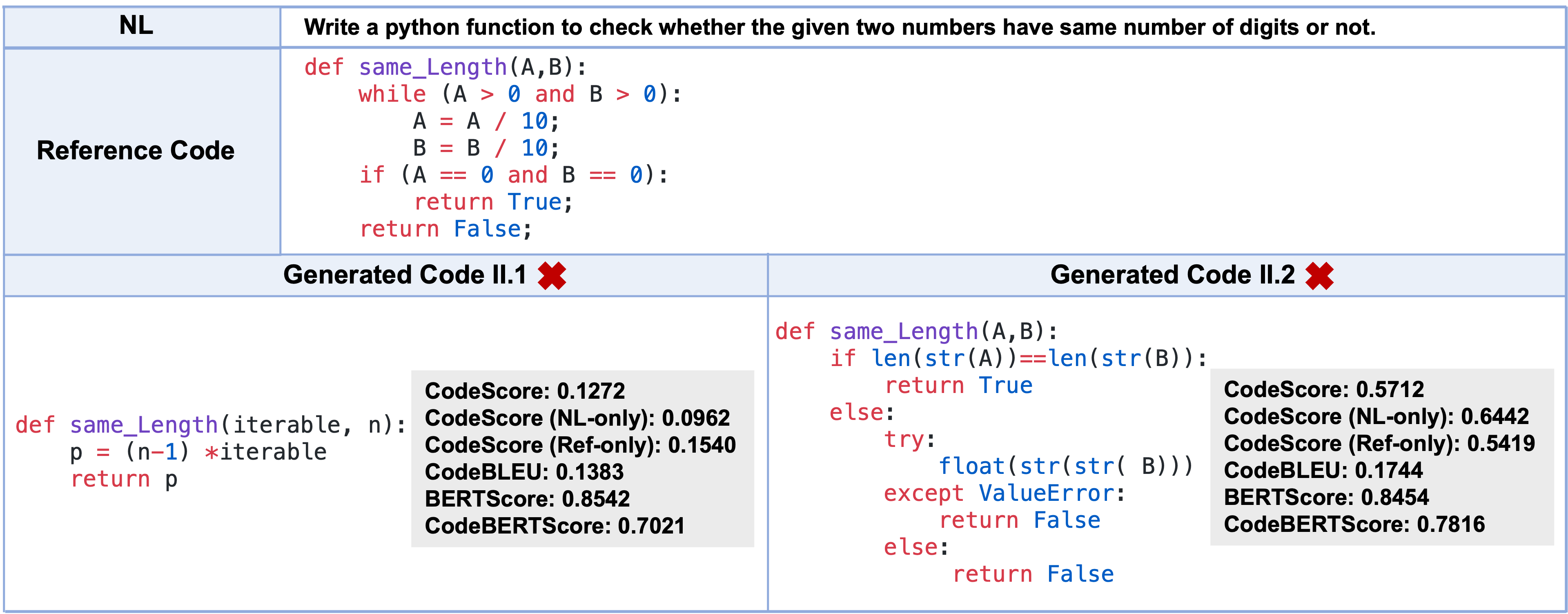}}\\
 	\subfloat[Case III]{\label{fig:c}\includegraphics[width=\textwidth]{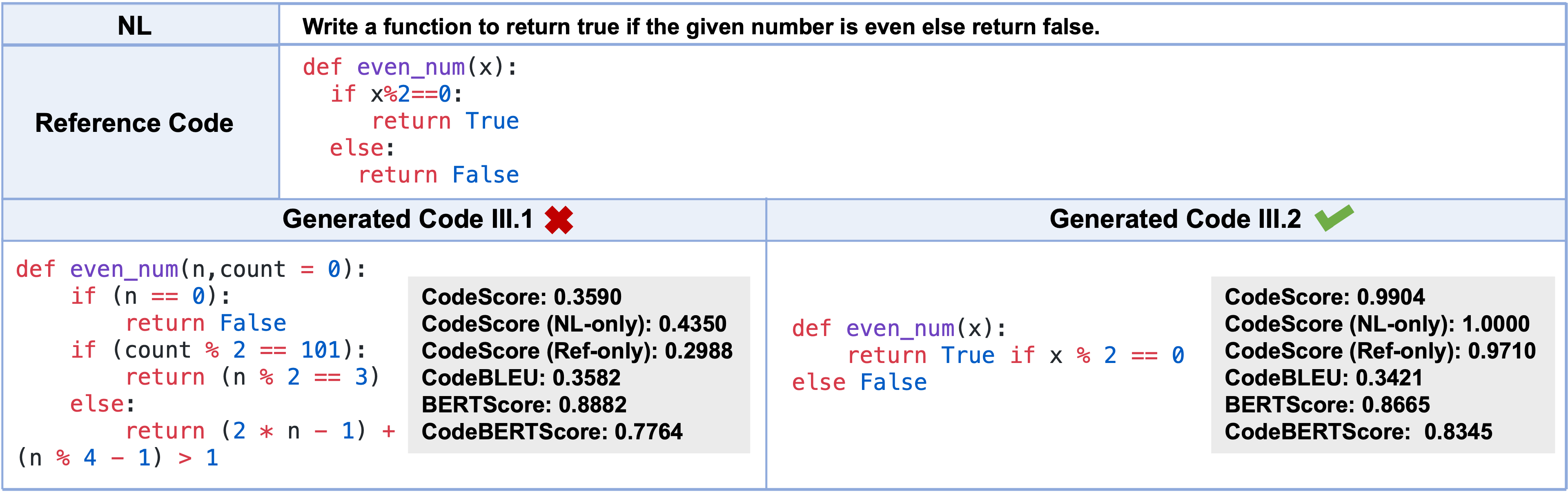}}\\
	\caption{Case study of different EMs. For each case, the second generated code is superior to the first one.}
        \label{fig:d}
\end{figure}

\begin{figure}[ht!]
    \centering
    \includegraphics[width=\textwidth]{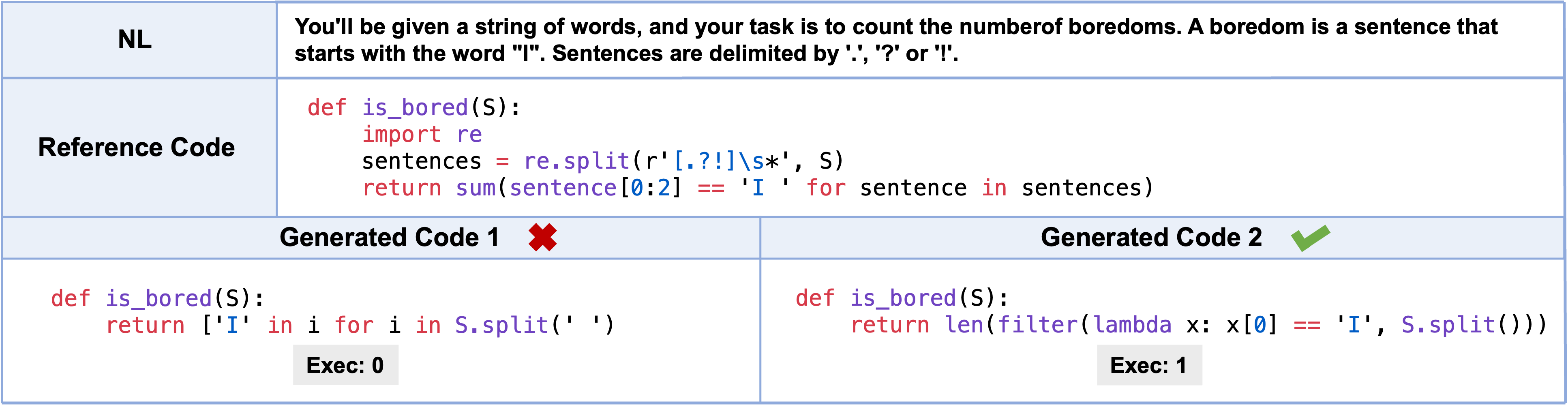}
    \caption{Case study of Exec.}
    \label{fig:case_exec}
\end{figure}

We present the results of the human evaluation in Table \ref{HECS}. Remarkably, our proposed CodeScore significantly outperforms all other EMs. Relative to these, CodeScore shows an improvement of at least 54.6\% in the human evaluation. All p-values are substantially less than 0.005 \footnote{\jx{The smaller the p-value, the less likely it is that the results are due to random factors.}}, underscoring that these improvements are statistically significant.

\vspace{6pt}
\RS{4}{Human evaluation indicates that CodeScore shows significant improvements over previous representative EMs.}

\subsubsection{RQ5: Case Study}
Fig. \ref{fig:d} displays a selection of generated codes and their corresponding EM scores (as per Section \ref{Human Evaluation}) on MBPP-Eval dataset. It becomes evident that CodeBLEU, BERTScore, and CodeBERTScore each exhibit unique issues. From these examples, we glean the following insights: 1) CodeBLEU tends to assign relatively low scores to generated code, even when the code is functionally correct. Furthermore, it appears to favor generated codes that maintain structural consistency with the reference code. For instance, even though Generated Code III.2 is functionally correct, it receives a lower CodeBLEU score than III.1, which is fundamentally incorrect. 2) Both BERTScore and CodeBERTScore have a propensity to award relatively high scores to generated code, even when the code is essentially flawed. Additionally, they often assign lower scores to better generated codes. For example, Generated Code II/III.2 has a lower BERTScore than II/III.1, and Generated Code I.2 has a lower CodeBERTScore than I.1. In contrast, CodeScore performs admirably in all of these scenarios. 
Our CodeScore aligns more closely with Ground Truth compared to other EMs.
Moreover, the various formats of input have little impact on CodeScore's scorings, indicating that CodeScore can effectively make judgments based on natural language and/or reference code, adapting to different input formats.

We further examine Exec's capabilities through a case study. We find that Exec can effectively discriminate the cases of successful and unsuccessful compilation, especially sensitive to some errors that lead to compilation failures. A representative example is shown in Figure \ref{fig:case_exec}, where in Generated Code 1, the code with mismatched parentheses is recognized by Exec, and in Generated Code 2, the code with multiple nested parentheses is not misidentified by Exec.

\vspace{6pt}
\RS{5}{Through case studies, we find that our approach does not have the problems faced by previous EMs and is effective in evaluating the functional correctness and compilability of generated code.}

\section{Threats to Validity}
There are two major threats to the validity of our work. 
1) Threats to external validity concern the quality of experimental datasets and the generalizability of our results. We evaluated our approach using three public code generation datasets, which are considered mainstream benchmarks in the field and have been utilized extensively in prior research \cite{wizardcoder,starcoder,dataset_previous_1,dataset_previous_2,dataset_previous_3,dataset_previous_4}. Given their widespread use, we believe that the findings derived from these datasets offer a reasonable degree of generalizability and could potentially extend to other datasets.
2) Threats to internal validity involve the impact of hyperparameters and instability characteristics of deep learning models. Deep learning models exhibit a certain sensitivity to hyperparameter settings. In our approach, we conduct a small-range grid search on hyper-parameters using a distinct validation subset. The same set of hyperparameters is consistently applied across all datasets and compared with various baselines, achieving favorable performance consistently. Even with the same hyper-parameters, deep learning models still encounter instability issues due to factors such as the random initialization of model parameters and the random shuffling of training data. Therefore, in our experiments, we run UniCE 5 times and report its average performance.
For fairness, we also run other LLM-based metrics five times with their public source code and provide the average performance.

\section{Discussion}
While we have demonstrated that CodeScore is an effective LLM-based metric for code evaluation, we acknowledge that it still has certain limitations.
\begin{itemize}
    \item First, learning code execution for code evaluation requires collecting a certain amount of data, including sufficient test cases, generated codes, reference codes, and NL descriptions. However, collecting this data is far less expensive than performing human evaluation.
    \item Second, in this paper, CodeScore is more suitable for evaluating function-level code in Python. Nevertheless, our work establishes the viability of code evaluation based on UniCE, and this approach can feasibly be extended to other scenarios. We aim to broaden CodeScore to encompass a wider range of codes in our future work. 
    \item Third, employing CodeScore for code evaluation entails additional computation and time. However, we maintain that this is still within an acceptable range, considering the benefits it provides in terms of the accuracy and reliability of code evaluation.
\end{itemize}

\section{Conclusion and Future Work}
\label{Conclusion and Discussion}
In this paper, we have proposed a code evaluation learning framework based on LLMs with a unified input, which we refer to as UniCE. UniCe is designed to learn the code execution of generated code. In response to the imprecise evaluations provided by existing match-based CEMs and LLM-based EMs, we introduced CodeScore based on UniCE, which is an effective CEM to measure the functional correctness of generated code. Furthermore, our CodeScore can be applied to three application scenarios (Ref-only, NL-only, and Ref\&NL) for code evaluation with a unified input. This is in contrast to traditional CEMs, which typically only consider the Ref-only scenario. To validate CodeScore, we constructed three code evaluation datasets (i.e., APPS-Eval, MBPP-Eval, and HE-Eval), which correspond to three popular benchmark datasets in code generation (i.e., MBPP, APPS, and HumanEval). Experimental results affirm the efficacy of CodeScore, which achieves state-of-the-art performance on multiple code evaluation datasets. 

We hope this work sheds light on future work in the direction of LLM-based code evaluation. Our code evaluation dataset can serve as a benchmark for evaluating the functional correctness of generated code. Furthermore, our work can be applied to facilitate the training of code generation models by providing positive feedback.

\begin{acks}
This research is supported by the National Natural Science Foundation of China under Grant No.62192733, 61832009, 62192731, 62192730, 62072007, the Key Program of Hubei under Grant JD2023008.
\end{acks}

\bibliographystyle{ACM-Reference-Format}
\bibliography{ref}


\begin{thebibliography}{68}


\ifx \showCODEN    \undefined \def \showCODEN     #1{\unskip}     \fi
\ifx \showDOI      \undefined \def \showDOI       #1{#1}\fi
\ifx \showISBNx    \undefined \def \showISBNx     #1{\unskip}     \fi
\ifx \showISBNxiii \undefined \def \showISBNxiii  #1{\unskip}     \fi
\ifx \showISSN     \undefined \def \showISSN      #1{\unskip}     \fi
\ifx \showLCCN     \undefined \def \showLCCN      #1{\unskip}     \fi
\ifx \shownote     \undefined \def \shownote      #1{#1}          \fi
\ifx \showarticletitle \undefined \def \showarticletitle #1{#1}   \fi
\ifx \showURL      \undefined \def \showURL       {\relax}        \fi
\providecommand\bibfield[2]{#2}
\providecommand\bibinfo[2]{#2}
\providecommand\natexlab[1]{#1}
\providecommand\showeprint[2][]{arXiv:#2}

\bibitem[Arakelyan et~al\mbox{.}(2022)]%
        {NS3}
\bibfield{author}{\bibinfo{person}{Shushan Arakelyan}, \bibinfo{person}{Anna
  Hakhverdyan}, \bibinfo{person}{Miltiadis Allamanis},
  \bibinfo{person}{Christophe Hauser}, \bibinfo{person}{Luis Garcia}, {and}
  \bibinfo{person}{Xiang Ren}.} \bibinfo{year}{2022}\natexlab{}.
\newblock \showarticletitle{{NS3:} Neuro-Symbolic Semantic Code Search}.
\newblock \bibinfo{journal}{\emph{CoRR}}  \bibinfo{volume}{abs/2205.10674}
  (\bibinfo{year}{2022}).
\newblock


\bibitem[Austin et~al\mbox{.}(2021)]%
        {MBPP}
\bibfield{author}{\bibinfo{person}{Jacob Austin}, \bibinfo{person}{Augustus
  Odena}, \bibinfo{person}{Maxwell~I. Nye}, \bibinfo{person}{Maarten Bosma},
  \bibinfo{person}{Henryk Michalewski}, \bibinfo{person}{David Dohan},
  \bibinfo{person}{Ellen Jiang}, \bibinfo{person}{Carrie~J. Cai},
  \bibinfo{person}{Michael Terry}, \bibinfo{person}{Quoc~V. Le}, {and}
  \bibinfo{person}{Charles Sutton}.} \bibinfo{year}{2021}\natexlab{}.
\newblock \showarticletitle{Program Synthesis with Large Language Models}.
\newblock \bibinfo{journal}{\emph{CoRR}}  \bibinfo{volume}{abs/2108.07732}
  (\bibinfo{year}{2021}).
\newblock


\bibitem[Banerjee and Lavie(2005)]%
        {METEOR}
\bibfield{author}{\bibinfo{person}{Satanjeev Banerjee} {and}
  \bibinfo{person}{Alon Lavie}.} \bibinfo{year}{2005}\natexlab{}.
\newblock \showarticletitle{{METEOR:} An Automatic Metric for {MT} Evaluation
  with Improved Correlation with Human Judgments}. In
  \bibinfo{booktitle}{\emph{IEEvaluation@ACL}}. \bibinfo{publisher}{Association
  for Computational Linguistics}, \bibinfo{pages}{65--72}.
\newblock


\bibitem[Bravais(1844)]%
        {bravais1844analyse}
\bibfield{author}{\bibinfo{person}{Auguste Bravais}.}
  \bibinfo{year}{1844}\natexlab{}.
\newblock \bibinfo{booktitle}{\emph{Analyse math{\'e}matique sur les
  probabilit{\'e}s des erreurs de situation d'un point}}.
\newblock \bibinfo{publisher}{Impr. Royale}.
\newblock


\bibitem[Chen et~al\mbox{.}(2021)]%
        {CodeX}
\bibfield{author}{\bibinfo{person}{Mark Chen}, \bibinfo{person}{Jerry Tworek},
  \bibinfo{person}{Heewoo Jun}, \bibinfo{person}{Qiming Yuan},
  \bibinfo{person}{Henrique~Pond{\'{e}} de Oliveira~Pinto},
  \bibinfo{person}{Jared Kaplan}, \bibinfo{person}{Harrison Edwards},
  \bibinfo{person}{Yuri Burda}, \bibinfo{person}{Nicholas Joseph},
  \bibinfo{person}{Greg Brockman}, \bibinfo{person}{Alex Ray},
  \bibinfo{person}{Raul Puri}, \bibinfo{person}{Gretchen Krueger},
  \bibinfo{person}{Michael Petrov}, \bibinfo{person}{Heidy Khlaaf},
  \bibinfo{person}{Girish Sastry}, \bibinfo{person}{Pamela Mishkin},
  \bibinfo{person}{Brooke Chan}, \bibinfo{person}{Scott Gray},
  \bibinfo{person}{Nick Ryder}, \bibinfo{person}{Mikhail Pavlov},
  \bibinfo{person}{Alethea Power}, \bibinfo{person}{Lukasz Kaiser},
  \bibinfo{person}{Mohammad Bavarian}, \bibinfo{person}{Clemens Winter},
  \bibinfo{person}{Philippe Tillet}, \bibinfo{person}{Felipe~Petroski Such},
  \bibinfo{person}{Dave Cummings}, \bibinfo{person}{Matthias Plappert},
  \bibinfo{person}{Fotios Chantzis}, \bibinfo{person}{Elizabeth Barnes},
  \bibinfo{person}{Ariel Herbert{-}Voss}, \bibinfo{person}{William~Hebgen
  Guss}, \bibinfo{person}{Alex Nichol}, \bibinfo{person}{Alex Paino},
  \bibinfo{person}{Nikolas Tezak}, \bibinfo{person}{Jie Tang},
  \bibinfo{person}{Igor Babuschkin}, \bibinfo{person}{Suchir Balaji},
  \bibinfo{person}{Shantanu Jain}, \bibinfo{person}{William Saunders},
  \bibinfo{person}{Christopher Hesse}, \bibinfo{person}{Andrew~N. Carr},
  \bibinfo{person}{Jan Leike}, \bibinfo{person}{Joshua Achiam},
  \bibinfo{person}{Vedant Misra}, \bibinfo{person}{Evan Morikawa},
  \bibinfo{person}{Alec Radford}, \bibinfo{person}{Matthew Knight},
  \bibinfo{person}{Miles Brundage}, \bibinfo{person}{Mira Murati},
  \bibinfo{person}{Katie Mayer}, \bibinfo{person}{Peter Welinder},
  \bibinfo{person}{Bob McGrew}, \bibinfo{person}{Dario Amodei},
  \bibinfo{person}{Sam McCandlish}, \bibinfo{person}{Ilya Sutskever}, {and}
  \bibinfo{person}{Wojciech Zaremba}.} \bibinfo{year}{2021}\natexlab{}.
\newblock \showarticletitle{Evaluating Large Language Models Trained on Code}.
\newblock \bibinfo{journal}{\emph{CoRR}} (\bibinfo{year}{2021}).
\newblock
\urldef\tempurl%
\url{https://arxiv.org/abs/2107.03374}
\showURL{%
\tempurl}


\bibitem[Devlin et~al\mbox{.}(2019)]%
        {BERT}
\bibfield{author}{\bibinfo{person}{Jacob Devlin}, \bibinfo{person}{Ming{-}Wei
  Chang}, \bibinfo{person}{Kenton Lee}, {and} \bibinfo{person}{Kristina
  Toutanova}.} \bibinfo{year}{2019}\natexlab{}.
\newblock \showarticletitle{{BERT:} Pre-training of Deep Bidirectional
  Transformers for Language Understanding}. In
  \bibinfo{booktitle}{\emph{{NAACL-HLT} {(1)}}}.
  \bibinfo{publisher}{Association for Computational Linguistics},
  \bibinfo{pages}{4171--4186}.
\newblock


\bibitem[Dong et~al\mbox{.}(2023a)]%
        {dongself}
\bibfield{author}{\bibinfo{person}{Yihong Dong}, \bibinfo{person}{Xue Jiang},
  \bibinfo{person}{Zhi Jin}, {and} \bibinfo{person}{Ge Li}.}
  \bibinfo{year}{2023}\natexlab{a}.
\newblock \showarticletitle{Self-collaboration Code Generation via ChatGPT}.
\newblock \bibinfo{journal}{\emph{ACM Transactions on Software Engineering and
  Methodology}} (\bibinfo{year}{2023}).
\newblock


\bibitem[Dong et~al\mbox{.}(2024a)]%
        {dong2024generalization}
\bibfield{author}{\bibinfo{person}{Yihong Dong}, \bibinfo{person}{Xue Jiang},
  \bibinfo{person}{Huanyu Liu}, \bibinfo{person}{Zhi Jin}, {and}
  \bibinfo{person}{Ge Li}.} \bibinfo{year}{2024}\natexlab{a}.
\newblock \showarticletitle{Generalization or memorization: Data contamination
  and trustworthy evaluation for large language models}.
\newblock \bibinfo{journal}{\emph{arXiv preprint arXiv:2402.15938}}
  (\bibinfo{year}{2024}).
\newblock


\bibitem[Dong et~al\mbox{.}(2023c)]%
        {dong2023antecedent}
\bibfield{author}{\bibinfo{person}{Yihong Dong}, \bibinfo{person}{Ge Li},
  \bibinfo{person}{Xue Jiang}, {and} \bibinfo{person}{Zhi Jin}.}
  \bibinfo{year}{2023}\natexlab{c}.
\newblock \showarticletitle{Antecedent Predictions Are More Important Than You
  Think: An Effective Method for Tree-Based Code Generation}.
\newblock In \bibinfo{booktitle}{\emph{ECAI 2023}}. \bibinfo{publisher}{IOS
  Press}, \bibinfo{pages}{565--574}.
\newblock


\bibitem[Dong et~al\mbox{.}(2023b)]%
        {CODEP}
\bibfield{author}{\bibinfo{person}{Yihong Dong}, \bibinfo{person}{Ge Li}, {and}
  \bibinfo{person}{Zhi Jin}.} \bibinfo{year}{2023}\natexlab{b}.
\newblock \showarticletitle{{CODEP:} Grammatical Seq2Seq Model for
  General-Purpose Code Generation}. In \bibinfo{booktitle}{\emph{{ISSTA}}}.
  \bibinfo{publisher}{{ACM}}, \bibinfo{pages}{188--198}.
\newblock


\bibitem[Dong et~al\mbox{.}(2024b)]%
        {dong-etal-2024-pace}
\bibfield{author}{\bibinfo{person}{Yihong Dong}, \bibinfo{person}{Kangcheng
  Luo}, \bibinfo{person}{Xue Jiang}, \bibinfo{person}{Zhi Jin}, {and}
  \bibinfo{person}{Ge Li}.} \bibinfo{year}{2024}\natexlab{b}.
\newblock \showarticletitle{{PACE}: Improving Prompt with Actor-Critic Editing
  for Large Language Model}. In \bibinfo{booktitle}{\emph{Findings of the
  Association for Computational Linguistics ACL 2024}}.
  \bibinfo{pages}{7304--7323}.
\newblock


\bibitem[Eghbali and Pradel(2022)]%
        {CrystalBLEU}
\bibfield{author}{\bibinfo{person}{Aryaz Eghbali} {and}
  \bibinfo{person}{Michael Pradel}.} \bibinfo{year}{2022}\natexlab{}.
\newblock \showarticletitle{CrystalBLEU: Precisely and Efficiently Measuring
  the Similarity of Code}. In \bibinfo{booktitle}{\emph{{ASE}}}.
  \bibinfo{publisher}{{ACM}}, \bibinfo{pages}{28:1--28:12}.
\newblock


\bibitem[Feng et~al\mbox{.}(2020)]%
        {CodeBERT}
\bibfield{author}{\bibinfo{person}{Zhangyin Feng}, \bibinfo{person}{Daya Guo},
  \bibinfo{person}{Duyu Tang}, \bibinfo{person}{Nan Duan},
  \bibinfo{person}{Xiaocheng Feng}, \bibinfo{person}{Ming Gong},
  \bibinfo{person}{Linjun Shou}, \bibinfo{person}{Bing Qin},
  \bibinfo{person}{Ting Liu}, \bibinfo{person}{Daxin Jiang}, {and}
  \bibinfo{person}{Ming Zhou}.} \bibinfo{year}{2020}\natexlab{}.
\newblock \showarticletitle{CodeBERT: {A} Pre-Trained Model for Programming and
  Natural Languages}. In \bibinfo{booktitle}{\emph{{EMNLP} (Findings)}}
  \emph{(\bibinfo{series}{Findings of {ACL}}, Vol.~\bibinfo{volume}{{EMNLP}
  2020})}. \bibinfo{publisher}{Association for Computational Linguistics},
  \bibinfo{pages}{1536--1547}.
\newblock


\bibitem[Fried et~al\mbox{.}(2022)]%
        {fried2022incoder}
\bibfield{author}{\bibinfo{person}{Daniel Fried}, \bibinfo{person}{Armen
  Aghajanyan}, \bibinfo{person}{Jessy Lin}, \bibinfo{person}{Sida Wang},
  \bibinfo{person}{Eric Wallace}, \bibinfo{person}{Freda Shi},
  \bibinfo{person}{Ruiqi Zhong}, \bibinfo{person}{Wen-tau Yih},
  \bibinfo{person}{Luke Zettlemoyer}, {and} \bibinfo{person}{Mike Lewis}.}
  \bibinfo{year}{2022}\natexlab{}.
\newblock \showarticletitle{Incoder: A generative model for code infilling and
  synthesis}.
\newblock \bibinfo{journal}{\emph{arXiv preprint arXiv:2204.05999}}
  (\bibinfo{year}{2022}).
\newblock


\bibitem[Guo et~al\mbox{.}(2022a)]%
        {UniXcoder}
\bibfield{author}{\bibinfo{person}{Daya Guo}, \bibinfo{person}{Shuai Lu},
  \bibinfo{person}{Nan Duan}, \bibinfo{person}{Yanlin Wang},
  \bibinfo{person}{Ming Zhou}, {and} \bibinfo{person}{Jian Yin}.}
  \bibinfo{year}{2022}\natexlab{a}.
\newblock \showarticletitle{UniXcoder: Unified Cross-Modal Pre-training for
  Code Representation}. In \bibinfo{booktitle}{\emph{{ACL} {(1)}}}.
  \bibinfo{publisher}{Association for Computational Linguistics},
  \bibinfo{pages}{7212--7225}.
\newblock


\bibitem[Guo et~al\mbox{.}(2022b)]%
        {GuoS0DBA22}
\bibfield{author}{\bibinfo{person}{Daya Guo}, \bibinfo{person}{Alexey
  Svyatkovskiy}, \bibinfo{person}{Jian Yin}, \bibinfo{person}{Nan Duan},
  \bibinfo{person}{Marc Brockschmidt}, {and} \bibinfo{person}{Miltiadis
  Allamanis}.} \bibinfo{year}{2022}\natexlab{b}.
\newblock \showarticletitle{Learning to Complete Code with Sketches}. In
  \bibinfo{booktitle}{\emph{ICLR}}. \bibinfo{publisher}{OpenReview.net}.
\newblock


\bibitem[Hao et~al\mbox{.}(2022)]%
        {AixBench}
\bibfield{author}{\bibinfo{person}{Yiyang Hao}, \bibinfo{person}{Ge Li},
  \bibinfo{person}{Yongqiang Liu}, \bibinfo{person}{Xiaowei Miao},
  \bibinfo{person}{He Zong}, \bibinfo{person}{Siyuan Jiang},
  \bibinfo{person}{Yang Liu}, {and} \bibinfo{person}{He Wei}.}
  \bibinfo{year}{2022}\natexlab{}.
\newblock \showarticletitle{AixBench: {A} Code Generation Benchmark Dataset}.
\newblock \bibinfo{journal}{\emph{CoRR}}  \bibinfo{volume}{abs/2206.13179}
  (\bibinfo{year}{2022}).
\newblock


\bibitem[Hendrycks et~al\mbox{.}(2021)]%
        {APPS}
\bibfield{author}{\bibinfo{person}{Dan Hendrycks}, \bibinfo{person}{Steven
  Basart}, \bibinfo{person}{Saurav Kadavath}, \bibinfo{person}{Mantas Mazeika},
  \bibinfo{person}{Akul Arora}, \bibinfo{person}{Ethan Guo},
  \bibinfo{person}{Collin Burns}, \bibinfo{person}{Samir Puranik},
  \bibinfo{person}{Horace He}, \bibinfo{person}{Dawn Song}, {and}
  \bibinfo{person}{Jacob Steinhardt}.} \bibinfo{year}{2021}\natexlab{}.
\newblock \showarticletitle{Measuring Coding Challenge Competence With {APPS}}.
  In \bibinfo{booktitle}{\emph{NeurIPS Datasets and Benchmarks}}.
\newblock


\bibitem[Huang et~al\mbox{.}(2023)]%
        {dataset_previous_2}
\bibfield{author}{\bibinfo{person}{Dong Huang}, \bibinfo{person}{Qingwen Bu},
  \bibinfo{person}{Jie Zhang}, \bibinfo{person}{Xiaofei Xie},
  \bibinfo{person}{Junjie Chen}, {and} \bibinfo{person}{Heming Cui}.}
  \bibinfo{year}{2023}\natexlab{}.
\newblock \showarticletitle{Bias assessment and mitigation in llm-based code
  generation}.
\newblock \bibinfo{journal}{\emph{arXiv preprint arXiv:2309.14345}}
  (\bibinfo{year}{2023}).
\newblock


\bibitem[Inala et~al\mbox{.}(2022)]%
        {dataset_previous_1}
\bibfield{author}{\bibinfo{person}{Jeevana~Priya Inala},
  \bibinfo{person}{Chenglong Wang}, \bibinfo{person}{Mei Yang},
  \bibinfo{person}{Andr{\'{e}}s Codas}, \bibinfo{person}{Mark
  Encarnaci{\'{o}}n}, \bibinfo{person}{Shuvendu~K. Lahiri},
  \bibinfo{person}{Madanlal Musuvathi}, {and} \bibinfo{person}{Jianfeng Gao}.}
  \bibinfo{year}{2022}\natexlab{}.
\newblock \showarticletitle{Fault-Aware Neural Code Rankers}. In
  \bibinfo{booktitle}{\emph{NeurIPS}}.
\newblock


\bibitem[Jiang et~al\mbox{.}(2024)]%
        {jiang2024seed}
\bibfield{author}{\bibinfo{person}{Xue Jiang}, \bibinfo{person}{Yihong Dong},
  \bibinfo{person}{Zhi Jin}, {and} \bibinfo{person}{Ge Li}.}
  \bibinfo{year}{2024}\natexlab{}.
\newblock \showarticletitle{SEED: Customize Large Language Models with
  Sample-Efficient Adaptation for Code Generation}.
\newblock \bibinfo{journal}{\emph{arXiv preprint arXiv:2403.00046}}
  (\bibinfo{year}{2024}).
\newblock


\bibitem[Jiang et~al\mbox{.}(2023)]%
        {jiang2023self}
\bibfield{author}{\bibinfo{person}{Xue Jiang}, \bibinfo{person}{Yihong Dong},
  \bibinfo{person}{Lecheng Wang}, \bibinfo{person}{Fang Zheng},
  \bibinfo{person}{Qiwei Shang}, \bibinfo{person}{Ge Li}, \bibinfo{person}{Zhi
  Jin}, {and} \bibinfo{person}{Wenpin Jiao}.} \bibinfo{year}{2023}\natexlab{}.
\newblock \showarticletitle{Self-planning Code Generation with Large Language
  Models}.
\newblock \bibinfo{journal}{\emph{ACM Transactions on Software Engineering and
  Methodology}} (\bibinfo{year}{2023}).
\newblock


\bibitem[Kendall(1938)]%
        {kendall1938new}
\bibfield{author}{\bibinfo{person}{Maurice~G Kendall}.}
  \bibinfo{year}{1938}\natexlab{}.
\newblock \showarticletitle{A new measure of rank correlation}.
\newblock \bibinfo{journal}{\emph{Biometrika}} \bibinfo{volume}{30},
  \bibinfo{number}{1/2} (\bibinfo{year}{1938}), \bibinfo{pages}{81--93}.
\newblock


\bibitem[Kingma and Ba(2015)]%
        {adam}
\bibfield{author}{\bibinfo{person}{Diederik~P. Kingma} {and}
  \bibinfo{person}{Jimmy Ba}.} \bibinfo{year}{2015}\natexlab{}.
\newblock \showarticletitle{Adam: {A} Method for Stochastic Optimization}. In
  \bibinfo{booktitle}{\emph{{ICLR}}}.
\newblock


\bibitem[Kulal et~al\mbox{.}(2019)]%
        {Pass@k}
\bibfield{author}{\bibinfo{person}{Sumith Kulal}, \bibinfo{person}{Panupong
  Pasupat}, \bibinfo{person}{Kartik Chandra}, \bibinfo{person}{Mina Lee},
  \bibinfo{person}{Oded Padon}, \bibinfo{person}{Alex Aiken}, {and}
  \bibinfo{person}{Percy Liang}.} \bibinfo{year}{2019}\natexlab{}.
\newblock \showarticletitle{SPoC: Search-based Pseudocode to Code}. In
  \bibinfo{booktitle}{\emph{NeurIPS}}. \bibinfo{pages}{11883--11894}.
\newblock


\bibitem[Li et~al\mbox{.}(2024a)]%
        {li2024evocodebench}
\bibfield{author}{\bibinfo{person}{Jia Li}, \bibinfo{person}{Ge Li},
  \bibinfo{person}{Xuanming Zhang}, \bibinfo{person}{Yihong Dong}, {and}
  \bibinfo{person}{Zhi Jin}.} \bibinfo{year}{2024}\natexlab{a}.
\newblock \showarticletitle{EvoCodeBench: An Evolving Code Generation Benchmark
  Aligned with Real-World Code Repositories}.
\newblock \bibinfo{journal}{\emph{arXiv preprint arXiv:2404.00599}}
  (\bibinfo{year}{2024}).
\newblock


\bibitem[Li et~al\mbox{.}(2024b)]%
        {li-etal-2024-deveval}
\bibfield{author}{\bibinfo{person}{Jia Li}, \bibinfo{person}{Ge Li},
  \bibinfo{person}{Yunfei Zhao}, \bibinfo{person}{Yongmin Li},
  \bibinfo{person}{Huanyu Liu}, \bibinfo{person}{Hao Zhu},
  \bibinfo{person}{Lecheng Wang}, \bibinfo{person}{Kaibo Liu},
  \bibinfo{person}{Zheng Fang}, \bibinfo{person}{Lanshen Wang},
  \bibinfo{person}{Jiazheng Ding}, \bibinfo{person}{Xuanming Zhang},
  \bibinfo{person}{Yuqi Zhu}, \bibinfo{person}{Yihong Dong},
  \bibinfo{person}{Zhi Jin}, \bibinfo{person}{Binhua Li}, \bibinfo{person}{Fei
  Huang}, \bibinfo{person}{Yongbin Li}, \bibinfo{person}{Bin Gu}, {and}
  \bibinfo{person}{Mengfei Yang}.} \bibinfo{year}{2024}\natexlab{b}.
\newblock \showarticletitle{{D}ev{E}val: A Manually-Annotated Code Generation
  Benchmark Aligned with Real-World Code Repositories}. In
  \bibinfo{booktitle}{\emph{Findings of the Association for Computational
  Linguistics ACL 2024}}. \bibinfo{pages}{3603--3614}.
\newblock


\bibitem[Li et~al\mbox{.}(2023)]%
        {starcoder}
\bibfield{author}{\bibinfo{person}{Raymond Li}, \bibinfo{person}{Loubna~Ben
  Allal}, \bibinfo{person}{Yangtian Zi}, \bibinfo{person}{Niklas Muennighoff},
  \bibinfo{person}{Denis Kocetkov}, \bibinfo{person}{Chenghao Mou},
  \bibinfo{person}{Marc Marone}, \bibinfo{person}{Christopher Akiki},
  \bibinfo{person}{Jia Li}, \bibinfo{person}{Jenny Chim}, \bibinfo{person}{Qian
  Liu}, \bibinfo{person}{Evgenii Zheltonozhskii}, \bibinfo{person}{Terry~Yue
  Zhuo}, \bibinfo{person}{Thomas Wang}, \bibinfo{person}{Olivier Dehaene},
  \bibinfo{person}{Mishig Davaadorj}, \bibinfo{person}{Joel Lamy{-}Poirier},
  \bibinfo{person}{Jo{\~{a}}o Monteiro}, \bibinfo{person}{Oleh Shliazhko},
  \bibinfo{person}{Nicolas Gontier}, \bibinfo{person}{Nicholas Meade},
  \bibinfo{person}{Armel Zebaze}, \bibinfo{person}{Ming{-}Ho Yee},
  \bibinfo{person}{Logesh~Kumar Umapathi}, \bibinfo{person}{Jian Zhu},
  \bibinfo{person}{Benjamin Lipkin}, \bibinfo{person}{Muhtasham Oblokulov},
  \bibinfo{person}{Zhiruo Wang}, \bibinfo{person}{Rudra~Murthy V},
  \bibinfo{person}{Jason Stillerman}, \bibinfo{person}{Siva~Sankalp Patel},
  \bibinfo{person}{Dmitry Abulkhanov}, \bibinfo{person}{Marco Zocca},
  \bibinfo{person}{Manan Dey}, \bibinfo{person}{Zhihan Zhang},
  \bibinfo{person}{Nour Moustafa{-}Fahmy}, \bibinfo{person}{Urvashi
  Bhattacharyya}, \bibinfo{person}{Wenhao Yu}, \bibinfo{person}{Swayam Singh},
  \bibinfo{person}{Sasha Luccioni}, \bibinfo{person}{Paulo Villegas},
  \bibinfo{person}{Maxim Kunakov}, \bibinfo{person}{Fedor Zhdanov},
  \bibinfo{person}{Manuel Romero}, \bibinfo{person}{Tony Lee},
  \bibinfo{person}{Nadav Timor}, \bibinfo{person}{Jennifer Ding},
  \bibinfo{person}{Claire Schlesinger}, \bibinfo{person}{Hailey Schoelkopf},
  \bibinfo{person}{Jan Ebert}, \bibinfo{person}{Tri Dao},
  \bibinfo{person}{Mayank Mishra}, \bibinfo{person}{Alex Gu},
  \bibinfo{person}{Jennifer Robinson}, \bibinfo{person}{Carolyn~Jane Anderson},
  \bibinfo{person}{Brendan Dolan{-}Gavitt}, \bibinfo{person}{Danish
  Contractor}, \bibinfo{person}{Siva Reddy}, \bibinfo{person}{Daniel Fried},
  \bibinfo{person}{Dzmitry Bahdanau}, \bibinfo{person}{Yacine Jernite},
  \bibinfo{person}{Carlos~Mu{\~{n}}oz Ferrandis}, \bibinfo{person}{Sean
  Hughes}, \bibinfo{person}{Thomas Wolf}, \bibinfo{person}{Arjun Guha},
  \bibinfo{person}{Leandro von Werra}, {and} \bibinfo{person}{Harm de Vries}.}
  \bibinfo{year}{2023}\natexlab{}.
\newblock \showarticletitle{StarCoder: may the source be with you!}
\newblock \bibinfo{journal}{\emph{CoRR}}  \bibinfo{volume}{abs/2305.06161}
  (\bibinfo{year}{2023}).
\newblock


\bibitem[Li et~al\mbox{.}(2022)]%
        {alphacode}
\bibfield{author}{\bibinfo{person}{Yujia Li}, \bibinfo{person}{David Choi},
  \bibinfo{person}{Junyoung Chung}, \bibinfo{person}{Nate Kushman},
  \bibinfo{person}{Julian Schrittwieser}, \bibinfo{person}{R{\'e}mi Leblond},
  \bibinfo{person}{Tom Eccles}, \bibinfo{person}{James Keeling},
  \bibinfo{person}{Felix Gimeno}, \bibinfo{person}{Agustin Dal~Lago},
  {et~al\mbox{.}}} \bibinfo{year}{2022}\natexlab{}.
\newblock \showarticletitle{Competition-level code generation with alphacode}.
\newblock \bibinfo{journal}{\emph{Science}} \bibinfo{volume}{378},
  \bibinfo{number}{6624} (\bibinfo{year}{2022}), \bibinfo{pages}{1092--1097}.
\newblock


\bibitem[Lin(2004)]%
        {ROUGE}
\bibfield{author}{\bibinfo{person}{Chin-Yew Lin}.}
  \bibinfo{year}{2004}\natexlab{}.
\newblock \showarticletitle{{ROUGE}: A Package for Automatic Evaluation of
  Summaries}. In \bibinfo{booktitle}{\emph{Text Summarization Branches Out}}.
  \bibinfo{publisher}{Association for Computational Linguistics},
  \bibinfo{pages}{74--81}.
\newblock


\bibitem[Ling et~al\mbox{.}(2016)]%
        {LingBGHKWS16}
\bibfield{author}{\bibinfo{person}{Wang Ling}, \bibinfo{person}{Phil Blunsom},
  \bibinfo{person}{Edward Grefenstette}, \bibinfo{person}{Karl~Moritz Hermann},
  \bibinfo{person}{Tom{\'{a}}s Kocisk{\'{y}}}, \bibinfo{person}{Fumin Wang},
  {and} \bibinfo{person}{Andrew~W. Senior}.} \bibinfo{year}{2016}\natexlab{}.
\newblock \showarticletitle{Latent Predictor Networks for Code Generation}. In
  \bibinfo{booktitle}{\emph{{ACL} {(1)}}}. \bibinfo{publisher}{The Association
  for Computer Linguistics}.
\newblock


\bibitem[Lu et~al\mbox{.}(2022)]%
        {LuDHGHS22}
\bibfield{author}{\bibinfo{person}{Shuai Lu}, \bibinfo{person}{Nan Duan},
  \bibinfo{person}{Hojae Han}, \bibinfo{person}{Daya Guo},
  \bibinfo{person}{Seung{-}won Hwang}, {and} \bibinfo{person}{Alexey
  Svyatkovskiy}.} \bibinfo{year}{2022}\natexlab{}.
\newblock \showarticletitle{ReACC: {A} Retrieval-Augmented Code Completion
  Framework}. In \bibinfo{booktitle}{\emph{ACL}}.
  \bibinfo{publisher}{Association for Computational Linguistics},
  \bibinfo{pages}{6227--6240}.
\newblock


\bibitem[Luo et~al\mbox{.}(2023)]%
        {wizardcoder}
\bibfield{author}{\bibinfo{person}{Ziyang Luo}, \bibinfo{person}{Can Xu},
  \bibinfo{person}{Pu Zhao}, \bibinfo{person}{Qingfeng Sun},
  \bibinfo{person}{Xiubo Geng}, \bibinfo{person}{Wenxiang Hu},
  \bibinfo{person}{Chongyang Tao}, \bibinfo{person}{Jing Ma},
  \bibinfo{person}{Qingwei Lin}, {and} \bibinfo{person}{Daxin Jiang}.}
  \bibinfo{year}{2023}\natexlab{}.
\newblock \showarticletitle{WizardCoder: Empowering Code Large Language Models
  with Evol-Instruct}.
\newblock \bibinfo{journal}{\emph{CoRR}}  \bibinfo{volume}{abs/2306.08568}
  (\bibinfo{year}{2023}).
\newblock


\bibitem[Mood(1950)]%
        {mood1950introduction}
\bibfield{author}{\bibinfo{person}{Alexander~McFarlane Mood}.}
  \bibinfo{year}{1950}\natexlab{}.
\newblock \showarticletitle{Introduction to the Theory of Statistics.}
\newblock  (\bibinfo{year}{1950}).
\newblock


\bibitem[Mukherjee et~al\mbox{.}(2021)]%
        {MukherjeeWCRCJ21}
\bibfield{author}{\bibinfo{person}{Rohan Mukherjee}, \bibinfo{person}{Yeming
  Wen}, \bibinfo{person}{Dipak Chaudhari}, \bibinfo{person}{Thomas~W. Reps},
  \bibinfo{person}{Swarat Chaudhuri}, {and} \bibinfo{person}{Christopher~M.
  Jermaine}.} \bibinfo{year}{2021}\natexlab{}.
\newblock \showarticletitle{Neural Program Generation Modulo Static Analysis}.
  In \bibinfo{booktitle}{\emph{NeurIPS}}. \bibinfo{pages}{18984--18996}.
\newblock


\bibitem[Nijkamp et~al\mbox{.}(2023)]%
        {codegen}
\bibfield{author}{\bibinfo{person}{Erik Nijkamp}, \bibinfo{person}{Bo Pang},
  \bibinfo{person}{Hiroaki Hayashi}, \bibinfo{person}{Lifu Tu},
  \bibinfo{person}{Huan Wang}, \bibinfo{person}{Yingbo Zhou},
  \bibinfo{person}{Silvio Savarese}, {and} \bibinfo{person}{Caiming Xiong}.}
  \bibinfo{year}{2023}\natexlab{}.
\newblock \showarticletitle{CodeGen: An Open Large Language Model for Code with
  Multi-Turn Program Synthesis}. In \bibinfo{booktitle}{\emph{{ICLR}}}.
  \bibinfo{publisher}{OpenReview.net}.
\newblock


\bibitem[OpenAI(2023a)]%
        {ChatGPT}
\bibfield{author}{\bibinfo{person}{OpenAI}.} \bibinfo{year}{2023}\natexlab{a}.
\newblock \bibinfo{booktitle}{\emph{{ChatGPT: Optimizing Language Models for
  Dialogue}}}.
\newblock
\urldef\tempurl%
\url{https://openai.com/blog/chatgpt/}
\showURL{%
\tempurl}


\bibitem[OpenAI(2023b)]%
        {GPT-4}
\bibfield{author}{\bibinfo{person}{OpenAI}.} \bibinfo{year}{2023}\natexlab{b}.
\newblock \showarticletitle{{GPT-4} Technical Report}.
\newblock \bibinfo{journal}{\emph{CoRR}}  \bibinfo{volume}{abs/2303.08774}
  (\bibinfo{year}{2023}).
\newblock


\bibitem[Papineni et~al\mbox{.}(2002)]%
        {Bleu}
\bibfield{author}{\bibinfo{person}{Kishore Papineni}, \bibinfo{person}{Salim
  Roukos}, \bibinfo{person}{Todd Ward}, {and} \bibinfo{person}{Wei{-}Jing
  Zhu}.} \bibinfo{year}{2002}\natexlab{}.
\newblock \showarticletitle{Bleu: a Method for Automatic Evaluation of Machine
  Translation}. In \bibinfo{booktitle}{\emph{{ACL}}}.
  \bibinfo{publisher}{{ACL}}, \bibinfo{pages}{311--318}.
\newblock


\bibitem[Peters et~al\mbox{.}(2018)]%
        {ELMO}
\bibfield{author}{\bibinfo{person}{Matthew~E. Peters}, \bibinfo{person}{Mark
  Neumann}, \bibinfo{person}{Mohit Iyyer}, \bibinfo{person}{Matt Gardner},
  \bibinfo{person}{Christopher Clark}, \bibinfo{person}{Kenton Lee}, {and}
  \bibinfo{person}{Luke Zettlemoyer}.} \bibinfo{year}{2018}\natexlab{}.
\newblock \showarticletitle{Deep Contextualized Word Representations}. In
  \bibinfo{booktitle}{\emph{{NAACL-HLT}}}. \bibinfo{publisher}{Association for
  Computational Linguistics}, \bibinfo{pages}{2227--2237}.
\newblock


\bibitem[Popovic(2015)]%
        {chrF}
\bibfield{author}{\bibinfo{person}{Maja Popovic}.}
  \bibinfo{year}{2015}\natexlab{}.
\newblock \showarticletitle{chrF: character n-gram F-score for automatic {MT}
  evaluation}. In \bibinfo{booktitle}{\emph{WMT@EMNLP}}.
  \bibinfo{publisher}{The Association for Computer Linguistics},
  \bibinfo{pages}{392--395}.
\newblock


\bibitem[Ranasinghe et~al\mbox{.}(2020)]%
        {TransQuest}
\bibfield{author}{\bibinfo{person}{Tharindu Ranasinghe},
  \bibinfo{person}{Constantin Orasan}, {and} \bibinfo{person}{Ruslan Mitkov}.}
  \bibinfo{year}{2020}\natexlab{}.
\newblock \showarticletitle{TransQuest: Translation Quality Estimation with
  Cross-lingual Transformers}. In \bibinfo{booktitle}{\emph{{COLING}}}.
  \bibinfo{publisher}{International Committee on Computational Linguistics},
  \bibinfo{pages}{5070--5081}.
\newblock


\bibitem[Raychev et~al\mbox{.}(2014)]%
        {RaychevVY14}
\bibfield{author}{\bibinfo{person}{Veselin Raychev}, \bibinfo{person}{Martin~T.
  Vechev}, {and} \bibinfo{person}{Eran Yahav}.}
  \bibinfo{year}{2014}\natexlab{}.
\newblock \showarticletitle{Code completion with statistical language models}.
  In \bibinfo{booktitle}{\emph{{PLDI}}}. \bibinfo{publisher}{{ACM}},
  \bibinfo{pages}{419--428}.
\newblock


\bibitem[Rei et~al\mbox{.}(2022)]%
        {COMET-22}
\bibfield{author}{\bibinfo{person}{Ricardo Rei}, \bibinfo{person}{Jos{\'e}~GC
  De~Souza}, \bibinfo{person}{Duarte Alves}, \bibinfo{person}{Chrysoula Zerva},
  \bibinfo{person}{Ana~C Farinha}, \bibinfo{person}{Taisiya Glushkova},
  \bibinfo{person}{Alon Lavie}, \bibinfo{person}{Luisa Coheur}, {and}
  \bibinfo{person}{Andr{\'e}~FT Martins}.} \bibinfo{year}{2022}\natexlab{}.
\newblock \showarticletitle{COMET-22: Unbabel-IST 2022 Submission for the
  Metrics Shared Task}. In \bibinfo{booktitle}{\emph{WMT@EMNLP}}.
  \bibinfo{publisher}{Association for Computational Linguistics},
  \bibinfo{pages}{578--585}.
\newblock


\bibitem[Rei et~al\mbox{.}(2021)]%
        {COMET-21}
\bibfield{author}{\bibinfo{person}{Ricardo Rei}, \bibinfo{person}{Ana~C.
  Farinha}, \bibinfo{person}{Chrysoula Zerva}, \bibinfo{person}{Daan van
  Stigt}, \bibinfo{person}{Craig Stewart}, \bibinfo{person}{Pedro~G. Ramos},
  \bibinfo{person}{Taisiya Glushkova}, \bibinfo{person}{Andr{\'{e}} F.~T.
  Martins}, {and} \bibinfo{person}{Alon Lavie}.}
  \bibinfo{year}{2021}\natexlab{}.
\newblock \showarticletitle{Are References Really Needed? Unbabel-IST 2021
  Submission for the Metrics Shared Task}. In
  \bibinfo{booktitle}{\emph{WMT@EMNLP}}. \bibinfo{publisher}{Association for
  Computational Linguistics}, \bibinfo{pages}{1030--1040}.
\newblock


\bibitem[Rei et~al\mbox{.}(2020)]%
        {COMET}
\bibfield{author}{\bibinfo{person}{Ricardo Rei}, \bibinfo{person}{Craig
  Stewart}, \bibinfo{person}{Ana~C. Farinha}, {and} \bibinfo{person}{Alon
  Lavie}.} \bibinfo{year}{2020}\natexlab{}.
\newblock \showarticletitle{{COMET:} {A} Neural Framework for {MT} Evaluation}.
  In \bibinfo{booktitle}{\emph{{EMNLP} {(1)}}}. \bibinfo{publisher}{Association
  for Computational Linguistics}, \bibinfo{pages}{2685--2702}.
\newblock


\bibitem[Reimers and Gurevych(2019)]%
        {Sentence-BERT}
\bibfield{author}{\bibinfo{person}{Nils Reimers} {and} \bibinfo{person}{Iryna
  Gurevych}.} \bibinfo{year}{2019}\natexlab{}.
\newblock \showarticletitle{Sentence-BERT: Sentence Embeddings using Siamese
  BERT-Networks}. In \bibinfo{booktitle}{\emph{{EMNLP/IJCNLP} {(1)}}}.
  \bibinfo{publisher}{Association for Computational Linguistics},
  \bibinfo{pages}{3980--3990}.
\newblock


\bibitem[Ren et~al\mbox{.}(2020)]%
        {CodeBLEU}
\bibfield{author}{\bibinfo{person}{Shuo Ren}, \bibinfo{person}{Daya Guo},
  \bibinfo{person}{Shuai Lu}, \bibinfo{person}{Long Zhou},
  \bibinfo{person}{Shujie Liu}, \bibinfo{person}{Duyu Tang},
  \bibinfo{person}{Neel Sundaresan}, \bibinfo{person}{Ming Zhou},
  \bibinfo{person}{Ambrosio Blanco}, {and} \bibinfo{person}{Shuai Ma}.}
  \bibinfo{year}{2020}\natexlab{}.
\newblock \showarticletitle{CodeBLEU: a Method for Automatic Evaluation of Code
  Synthesis}.
\newblock \bibinfo{journal}{\emph{CoRR}}  \bibinfo{volume}{abs/2009.10297}
  (\bibinfo{year}{2020}).
\newblock


\bibitem[Rozi{\`{e}}re et~al\mbox{.}(2023)]%
        {codellama}
\bibfield{author}{\bibinfo{person}{Baptiste Rozi{\`{e}}re},
  \bibinfo{person}{Jonas Gehring}, \bibinfo{person}{Fabian Gloeckle},
  \bibinfo{person}{Sten Sootla}, \bibinfo{person}{Itai Gat},
  \bibinfo{person}{Xiaoqing~Ellen Tan}, \bibinfo{person}{Yossi Adi},
  \bibinfo{person}{Jingyu Liu}, \bibinfo{person}{Tal Remez},
  \bibinfo{person}{J{\'{e}}r{\'{e}}my Rapin}, \bibinfo{person}{Artyom
  Kozhevnikov}, \bibinfo{person}{Ivan Evtimov}, \bibinfo{person}{Joanna
  Bitton}, \bibinfo{person}{Manish Bhatt}, \bibinfo{person}{Cristian
  Canton{-}Ferrer}, \bibinfo{person}{Aaron Grattafiori},
  \bibinfo{person}{Wenhan Xiong}, \bibinfo{person}{Alexandre D{\'{e}}fossez},
  \bibinfo{person}{Jade Copet}, \bibinfo{person}{Faisal Azhar},
  \bibinfo{person}{Hugo Touvron}, \bibinfo{person}{Louis Martin},
  \bibinfo{person}{Nicolas Usunier}, \bibinfo{person}{Thomas Scialom}, {and}
  \bibinfo{person}{Gabriel Synnaeve}.} \bibinfo{year}{2023}\natexlab{}.
\newblock \showarticletitle{Code Llama: Open Foundation Models for Code}.
\newblock \bibinfo{journal}{\emph{CoRR}}  \bibinfo{volume}{abs/2308.12950}
  (\bibinfo{year}{2023}).
\newblock


\bibitem[Rozi{\`{e}}re et~al\mbox{.}(2020)]%
        {RoziereLCL20}
\bibfield{author}{\bibinfo{person}{Baptiste Rozi{\`{e}}re},
  \bibinfo{person}{Marie{-}Anne Lachaux}, \bibinfo{person}{Lowik Chanussot},
  {and} \bibinfo{person}{Guillaume Lample}.} \bibinfo{year}{2020}\natexlab{}.
\newblock \showarticletitle{Unsupervised Translation of Programming Languages}.
  In \bibinfo{booktitle}{\emph{NeurIPS}}.
\newblock


\bibitem[Sellam et~al\mbox{.}(2020)]%
        {BLEURT}
\bibfield{author}{\bibinfo{person}{Thibault Sellam}, \bibinfo{person}{Dipanjan
  Das}, {and} \bibinfo{person}{Ankur~P. Parikh}.}
  \bibinfo{year}{2020}\natexlab{}.
\newblock \showarticletitle{{BLEURT:} Learning Robust Metrics for Text
  Generation}. In \bibinfo{booktitle}{\emph{{ACL}}}.
  \bibinfo{publisher}{Association for Computational Linguistics},
  \bibinfo{pages}{7881--7892}.
\newblock


\bibitem[Shen et~al\mbox{.}(2022)]%
        {Industry}
\bibfield{author}{\bibinfo{person}{Sijie Shen}, \bibinfo{person}{Xiang Zhu},
  \bibinfo{person}{Yihong Dong}, \bibinfo{person}{Qizhi Guo},
  \bibinfo{person}{Yankun Zhen}, {and} \bibinfo{person}{Ge Li}.}
  \bibinfo{year}{2022}\natexlab{}.
\newblock \showarticletitle{Incorporating domain knowledge through task
  augmentation for front-end JavaScript code generation}. In
  \bibinfo{booktitle}{\emph{{ESEC/SIGSOFT} {FSE}}}. \bibinfo{publisher}{{ACM}},
  \bibinfo{pages}{1533--1543}.
\newblock


\bibitem[Sun et~al\mbox{.}(2022)]%
        {SunFCTHZ22}
\bibfield{author}{\bibinfo{person}{Weisong Sun}, \bibinfo{person}{Chunrong
  Fang}, \bibinfo{person}{Yuchen Chen}, \bibinfo{person}{Guanhong Tao},
  \bibinfo{person}{Tingxu Han}, {and} \bibinfo{person}{Quanjun Zhang}.}
  \bibinfo{year}{2022}\natexlab{}.
\newblock \showarticletitle{Code Search based on Context-aware Code
  Translation}. In \bibinfo{booktitle}{\emph{ICSE}}.
  \bibinfo{publisher}{{ACM}}, \bibinfo{pages}{388--400}.
\newblock


\bibitem[Sun et~al\mbox{.}(2020)]%
        {SunZXSMZ20}
\bibfield{author}{\bibinfo{person}{Zeyu Sun}, \bibinfo{person}{Qihao Zhu},
  \bibinfo{person}{Yingfei Xiong}, \bibinfo{person}{Yican Sun},
  \bibinfo{person}{Lili Mou}, {and} \bibinfo{person}{Lu Zhang}.}
  \bibinfo{year}{2020}\natexlab{}.
\newblock \showarticletitle{TreeGen: {A} Tree-Based Transformer Architecture
  for Code Generation}. In \bibinfo{booktitle}{\emph{{AAAI}}}.
  \bibinfo{publisher}{{AAAI} Press}, \bibinfo{pages}{8984--8991}.
\newblock


\bibitem[Tenney et~al\mbox{.}(2019)]%
        {TenneyDP19}
\bibfield{author}{\bibinfo{person}{Ian Tenney}, \bibinfo{person}{Dipanjan Das},
  {and} \bibinfo{person}{Ellie Pavlick}.} \bibinfo{year}{2019}\natexlab{}.
\newblock \showarticletitle{{BERT} Rediscovers the Classical {NLP} Pipeline}.
  In \bibinfo{booktitle}{\emph{{ACL} {(1)}}}. \bibinfo{publisher}{Association
  for Computational Linguistics}, \bibinfo{pages}{4593--4601}.
\newblock


\bibitem[Wan et~al\mbox{.}(2022)]%
        {UniTE}
\bibfield{author}{\bibinfo{person}{Yu Wan}, \bibinfo{person}{Dayiheng Liu},
  \bibinfo{person}{Baosong Yang}, \bibinfo{person}{Haibo Zhang},
  \bibinfo{person}{Boxing Chen}, \bibinfo{person}{Derek~F. Wong}, {and}
  \bibinfo{person}{Lidia~S. Chao}.} \bibinfo{year}{2022}\natexlab{}.
\newblock \showarticletitle{UniTE: Unified Translation Evaluation}. In
  \bibinfo{booktitle}{\emph{{ACL} {(1)}}}. \bibinfo{publisher}{Association for
  Computational Linguistics}, \bibinfo{pages}{8117--8127}.
\newblock


\bibitem[Wang et~al\mbox{.}(2021)]%
        {CodeT5}
\bibfield{author}{\bibinfo{person}{Yue Wang}, \bibinfo{person}{Weishi Wang},
  \bibinfo{person}{Shafiq~R. Joty}, {and} \bibinfo{person}{teven C.~H.~Hoi}.}
  \bibinfo{year}{2021}\natexlab{}.
\newblock \showarticletitle{CodeT5: Identifier-aware Unified Pre-trained
  Encoder-Decoder Models for Code Understanding and Generation}. In
  \bibinfo{booktitle}{\emph{{EMNLP} {(1)}}}. \bibinfo{pages}{8696--8708}.
\newblock


\bibitem[Wei et~al\mbox{.}(2019)]%
        {WeiBolin}
\bibfield{author}{\bibinfo{person}{Bolin Wei}, \bibinfo{person}{Ge Li},
  \bibinfo{person}{Xin Xia}, \bibinfo{person}{Zhiyi Fu}, {and}
  \bibinfo{person}{Zhi Jin}.} \bibinfo{year}{2019}\natexlab{}.
\newblock \showarticletitle{Code Generation as a Dual Task of Code
  Summarization}. In \bibinfo{booktitle}{\emph{NeurIPS}}.
  \bibinfo{pages}{6559--6569}.
\newblock


\bibitem[Wei et~al\mbox{.}(2023)]%
        {dataset_previous_4}
\bibfield{author}{\bibinfo{person}{Xiaokai Wei}, \bibinfo{person}{Sujan~Kumar
  Gonugondla}, \bibinfo{person}{Shiqi Wang}, \bibinfo{person}{Wasi~Uddin
  Ahmad}, \bibinfo{person}{Baishakhi Ray}, \bibinfo{person}{Haifeng Qian},
  \bibinfo{person}{Xiaopeng Li}, \bibinfo{person}{Varun Kumar},
  \bibinfo{person}{Zijian Wang}, \bibinfo{person}{Yuchen Tian},
  \bibinfo{person}{Qing Sun}, \bibinfo{person}{Ben Athiwaratkun},
  \bibinfo{person}{Mingyue Shang}, \bibinfo{person}{Murali~Krishna Ramanathan},
  \bibinfo{person}{Parminder Bhatia}, {and} \bibinfo{person}{Bing Xiang}.}
  \bibinfo{year}{2023}\natexlab{}.
\newblock \showarticletitle{Towards Greener Yet Powerful Code Generation via
  Quantization: An Empirical Study}. In
  \bibinfo{booktitle}{\emph{{ESEC/SIGSOFT} {FSE}}}. \bibinfo{publisher}{{ACM}},
  \bibinfo{pages}{224--236}.
\newblock


\bibitem[Yin and Neubig(2018)]%
        {YinN18}
\bibfield{author}{\bibinfo{person}{Pengcheng Yin} {and} \bibinfo{person}{Graham
  Neubig}.} \bibinfo{year}{2018}\natexlab{}.
\newblock \showarticletitle{{TRANX:} {A} Transition-based Neural Abstract
  Syntax Parser for Semantic Parsing and Code Generation}. In
  \bibinfo{booktitle}{\emph{EMNLP}}. \bibinfo{pages}{7--12}.
\newblock


\bibitem[Yuan et~al\mbox{.}(2021)]%
        {BARTScore}
\bibfield{author}{\bibinfo{person}{Weizhe Yuan}, \bibinfo{person}{Graham
  Neubig}, {and} \bibinfo{person}{Pengfei Liu}.}
  \bibinfo{year}{2021}\natexlab{}.
\newblock \showarticletitle{BARTScore: Evaluating Generated Text as Text
  Generation}. In \bibinfo{booktitle}{\emph{NeurIPS}}.
  \bibinfo{pages}{27263--27277}.
\newblock


\bibitem[Zhang et~al\mbox{.}(2023)]%
        {dataset_previous_3}
\bibfield{author}{\bibinfo{person}{Shun Zhang}, \bibinfo{person}{Zhenfang
  Chen}, \bibinfo{person}{Yikang Shen}, \bibinfo{person}{Mingyu Ding},
  \bibinfo{person}{Joshua~B. Tenenbaum}, {and} \bibinfo{person}{Chuang Gan}.}
  \bibinfo{year}{2023}\natexlab{}.
\newblock \showarticletitle{Planning with Large Language Models for Code
  Generation}. In \bibinfo{booktitle}{\emph{{ICLR}}}.
  \bibinfo{publisher}{OpenReview.net}.
\newblock


\bibitem[Zhang et~al\mbox{.}(2020)]%
        {BERTScore}
\bibfield{author}{\bibinfo{person}{Tianyi Zhang}, \bibinfo{person}{Varsha
  Kishore}, \bibinfo{person}{Felix Wu}, \bibinfo{person}{Kilian~Q. Weinberger},
  {and} \bibinfo{person}{Yoav Artzi}.} \bibinfo{year}{2020}\natexlab{}.
\newblock \showarticletitle{BERTScore: Evaluating Text Generation with {BERT}}.
  In \bibinfo{booktitle}{\emph{{ICLR}}}. \bibinfo{publisher}{OpenReview.net}.
\newblock


\bibitem[Zhao et~al\mbox{.}(2019)]%
        {MoverScore}
\bibfield{author}{\bibinfo{person}{Wei Zhao}, \bibinfo{person}{Maxime Peyrard},
  \bibinfo{person}{Fei Liu}, \bibinfo{person}{Yang Gao},
  \bibinfo{person}{Christian~M. Meyer}, {and} \bibinfo{person}{Steffen Eger}.}
  \bibinfo{year}{2019}\natexlab{}.
\newblock \showarticletitle{MoverScore: Text Generation Evaluating with
  Contextualized Embeddings and Earth Mover Distance}. In
  \bibinfo{booktitle}{\emph{{EMNLP/IJCNLP} {(1)}}}.
  \bibinfo{publisher}{Association for Computational Linguistics},
  \bibinfo{pages}{563--578}.
\newblock


\bibitem[Zhao et~al\mbox{.}(2023)]%
        {zhao2023seq2seq}
\bibfield{author}{\bibinfo{person}{Yunfei Zhao}, \bibinfo{person}{Yihong Dong},
  {and} \bibinfo{person}{Ge Li}.} \bibinfo{year}{2023}\natexlab{}.
\newblock \showarticletitle{Seq2Seq or Seq2Tree: Generating Code Using Both
  Paradigms via Mutual Learning}. In \bibinfo{booktitle}{\emph{Proceedings of
  the 14th Asia-Pacific Symposium on Internetware}}. \bibinfo{pages}{238--248}.
\newblock


\bibitem[Zheng et~al\mbox{.}(2023)]%
        {codegeex}
\bibfield{author}{\bibinfo{person}{Qinkai Zheng}, \bibinfo{person}{Xiao Xia},
  \bibinfo{person}{Xu Zou}, \bibinfo{person}{Yuxiao Dong},
  \bibinfo{person}{Shan Wang}, \bibinfo{person}{Yufei Xue},
  \bibinfo{person}{Zihan Wang}, \bibinfo{person}{Lei Shen},
  \bibinfo{person}{Andi Wang}, \bibinfo{person}{Yang Li}, \bibinfo{person}{Teng
  Su}, \bibinfo{person}{Zhilin Yang}, {and} \bibinfo{person}{Jie Tang}.}
  \bibinfo{year}{2023}\natexlab{}.
\newblock \showarticletitle{CodeGeeX: {A} Pre-Trained Model for Code Generation
  with Multilingual Evaluations on HumanEval-X}.
\newblock \bibinfo{journal}{\emph{CoRR}}  \bibinfo{volume}{abs/2303.17568}
  (\bibinfo{year}{2023}).
\newblock


\bibitem[Zhou et~al\mbox{.}(2023)]%
        {CodeBERTScore}
\bibfield{author}{\bibinfo{person}{Shuyan Zhou}, \bibinfo{person}{Uri Alon},
  \bibinfo{person}{Sumit Agarwal}, {and} \bibinfo{person}{Graham Neubig}.}
  \bibinfo{year}{2023}\natexlab{}.
\newblock \showarticletitle{CodeBERTScore: Evaluating Code Generation with
  Pretrained Models of Code}.
\newblock \bibinfo{journal}{\emph{CoRR}}  \bibinfo{volume}{abs/2302.05527}
  (\bibinfo{year}{2023}).
\newblock


\bibitem[Zhu et~al\mbox{.}(2022)]%
        {Zhu0R22}
\bibfield{author}{\bibinfo{person}{Ming Zhu}, \bibinfo{person}{Karthik Suresh},
  {and} \bibinfo{person}{Chandan~K. Reddy}.} \bibinfo{year}{2022}\natexlab{}.
\newblock \showarticletitle{Multilingual Code Snippets Training for Program
  Translation}. In \bibinfo{booktitle}{\emph{AAAI}}. \bibinfo{publisher}{{AAAI}
  Press}, \bibinfo{pages}{11783--11790}.
\newblock


\end{thebibliography}

\newpage
\appendix
\section{Test case generation via ChatGPT}
\label{Prompt of ChatGPT}
\begin{figure}[h!]
	\centering
\includegraphics[width=0.8\textwidth]{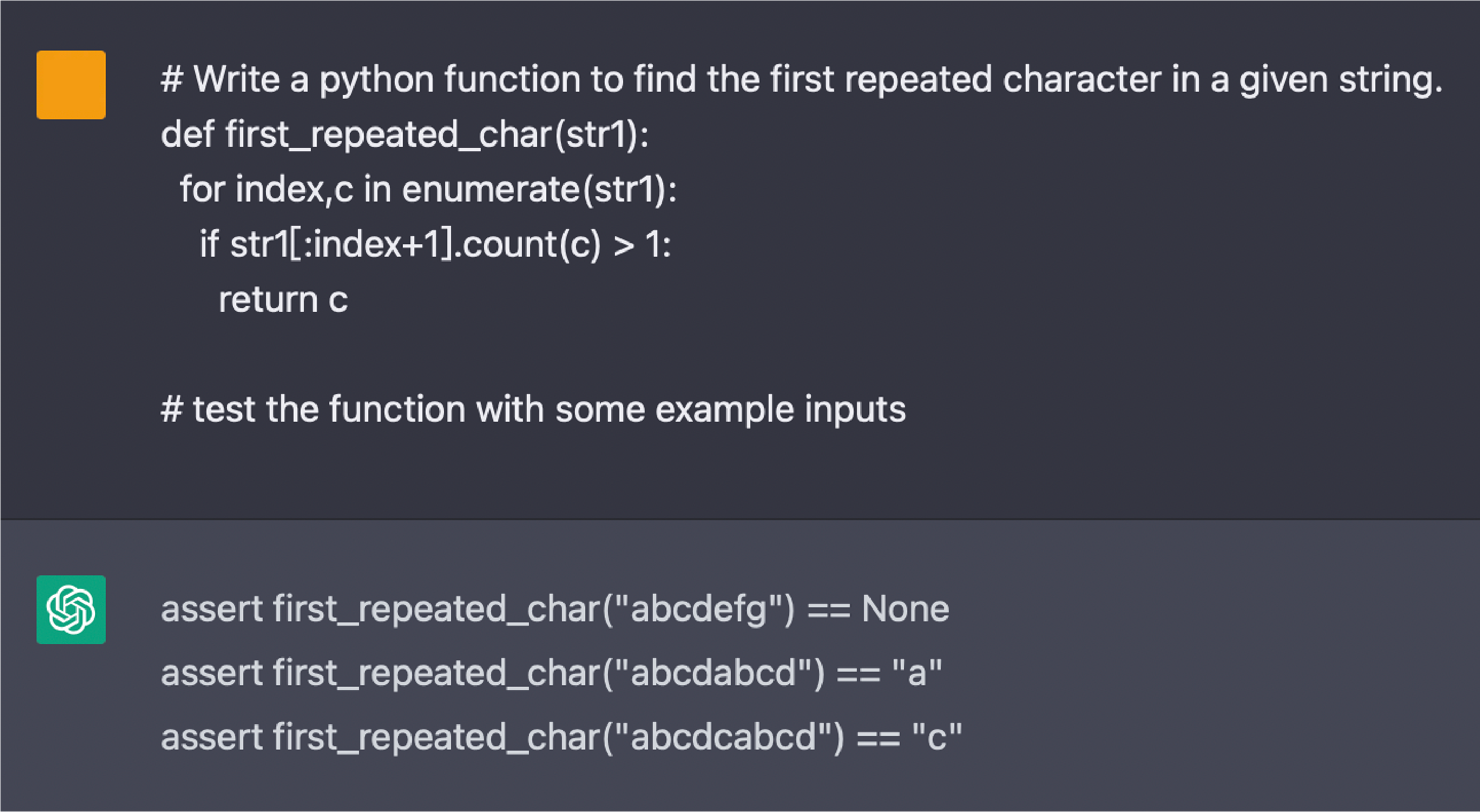}
	\caption{Example of ChatGPT generating test cases.}
	\label{Example of Test Case Generation}
\end{figure}

We randomly select 100 code generation tasks from the MBPP dataset and use the NL description and reference code of tasks to generate test cases via ChatGPT \cite{ChatGPT}. Fig. \ref{Example of Test Case Generation} shows an example of ChatGPT generating test cases. ChatGPT generates an average of 1.53 test cases per task.

\begin{figure}[h!]
	\centering
\includegraphics[width=0.75\textwidth]{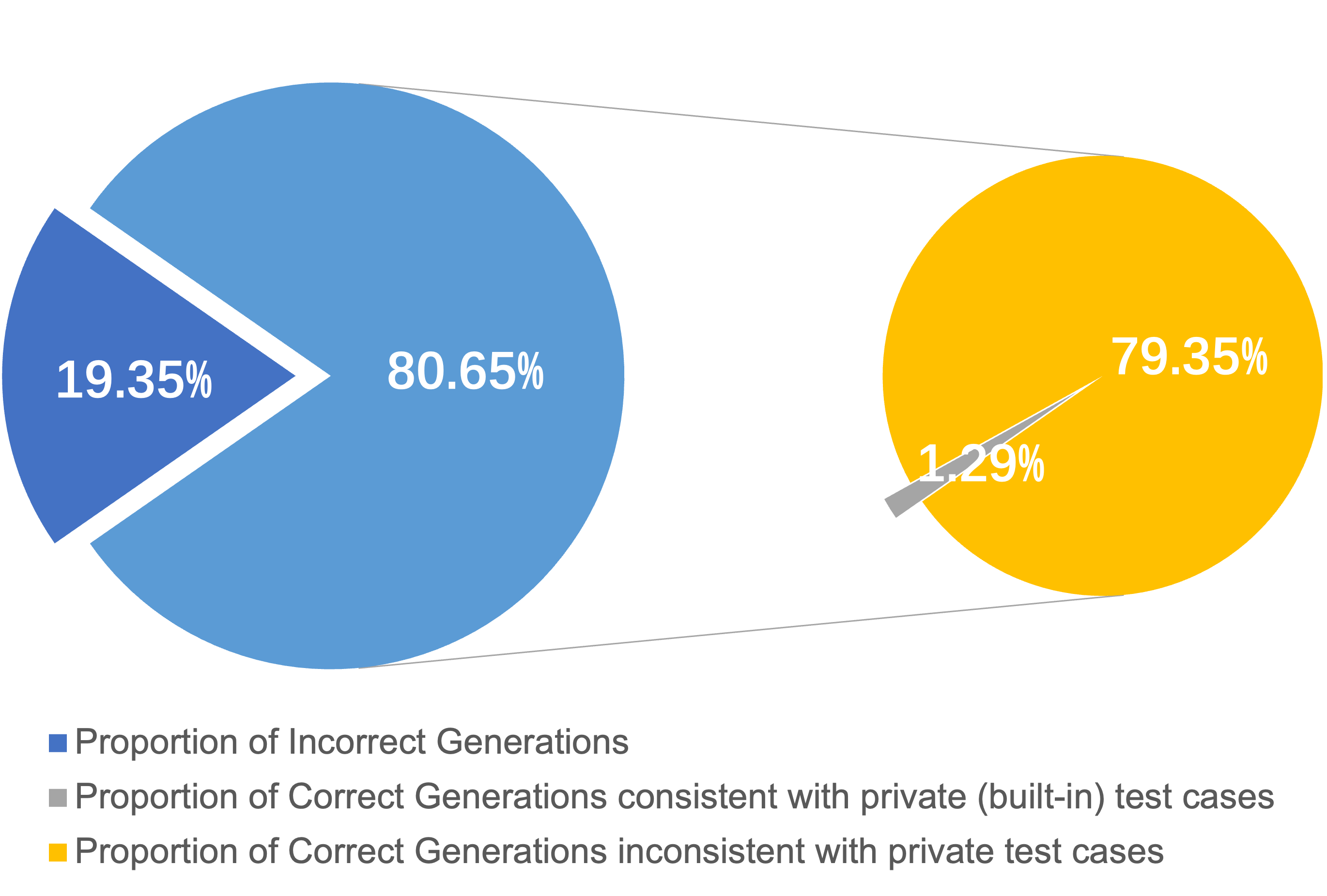}
	\caption{Test case generation via ChatGPT \cite{ChatGPT} in zero-shot setting (details can be found in Appendix \ref{Prompt of ChatGPT}).}
	\label{Test Case Generation}
\end{figure}

The results shown in Fig. \ref{Test Case Generation} indicate that \textbf{LLMs have the potential to judge the functional correctness of most programs with appropriate guidance}. Only 1.29\% Generations consistent with private test cases means that ChatGPT generates test cases by itself instead of copying private test cases. 

\end{document}